\documentclass[fleqn,usenatbib]{mnras}
\usepackage [latin1]{inputenc}
\usepackage{newtxtext,newtxmath}

\usepackage[T1]{fontenc}
\usepackage{ae,aecompl}

\usepackage{graphicx} 
\usepackage{amsmath}  
\usepackage{amssymb}  
\usepackage{hyperref}
\usepackage{longtable}
\usepackage{newtxtext,newtxmath}
\usepackage{booktabs}
\usepackage{morefloats}
\usepackage{footnote}
\usepackage{multirow}
\usepackage{color}
\definecolor{referee}{rgb}{0,0,0}
\definecolor{referee2}{rgb}{0,0,0}
\usepackage{pdflscape}
\usepackage{ragged2e}

\title[Redshift Searches of HerBS Sources]{IRAM 30m-EMIR Redshift Search of $z=3-4$ Lensed Dusty Starbursts selected from the HerBS sample}

\author[T. J. L. C. Bakx]{T. J. L. C. Bakx$^{1,2,3}$,\thanks{E-mail: bakx@a.phys.nagoya-u.ac.jp (Nagoya University)}
H. Dannerbauer$^{4,5}$,
D. Frayer$^{6}$,
S. A. Eales$^{1}$,
I. P\'erez-Fournon$^{4,5}$, \newauthor
Z.-Y. Cai$^{7,8}$,
D.L. Clements$^{9}$,
G. De Zotti$^{10}$,
J. Gonz{\'a}lez-Nuevo$^{11,12}$,
R.J. Ivison$^{13}$,\newauthor
A. Lapi$^{14,15}$,
M.~J.~Micha{\l}owski$^{16}$,
M. Negrello$^{1}$,
S. Serjeant$^{17}$,
M.W.L. Smith$^{1}$,\newauthor
P. Temi$^{18}$,
S. Urquhart$^{17}$,
P. van der Werf$^{19}$
\\
$^{1}$Department of Physics and Astronomy, Cardiff University, The Parade, Cardiff, CF24 3AA, United Kingdom\\
$^{2}$Division of Particle and Astrophysical Science, Graduate School of Science, Nagoya University, Aichi 464-8602, Japan.\\
$^{3}$National Astronomical Observatory of Japan, 2-21-1, Osawa, Mitaka, Tokyo 181-8588, Japan.\\
$^{4}$Instituto de Astrof\'isica de Canarias (IAC), E-38205 La Laguna, Tenerife, Spain\\
$^{5}$Universidad de La Laguna, Dpto. Astrof{\'i}sica, E-38206 La Laguna, Tenerife, Spain\\
$^{6}$Green Bank Observatory, PO Box 2, Green Bank, WV, 24944, USA\\
$^{7}$CAS Key Laboratory for Research in Galaxies and Cosmology, Department of Astronomy, \\ \hspace{0.13cm}University of Science and Technology of China, Hefei 230026, China\\
$^{8}$School of Astronomy and Space Science, University of Science and Technology of China, Hefei 230026, China\\
$^{9}$Blackett Lab, Imperial College, London, Prince Consort Road, London SW7 2AZ, UK \\
$^{10}$INAF-Osservatorio Astronomico di Padova, Vicolo dell'Osservatorio 5, I-35122 Padova, Italy\\
$^{11}$Departamento de F{\'i}sica, Universidad de Oviedo, C. Federico Garc{\'i}a Lorca 18, 33007 Oviedo, Spain\\
$^{12}$Instituto Universitario de Ciencias y Tecnolog{\'i}as Espaciales de Asturias (ICTEA), C. Independencia 13, 33004 Oviedo, Spain\\
$^{13}$European Southern Observatory, Karl-Schwarzschild-Strasse 2, D-85748 Garching, Germany \\
$^{14}$SISSA, Via Bonomea 265, I-34136 Trieste, Italy\\
$^{15}$IFPU - Institute for fundamental physics of the Universe, Via Beirut 2, 34014 Trieste, Italy\\
$^{16}$Astronomical Observatory Institute, Faculty of Physics, Adam Mickiewicz University, ul.~S{\l}oneczna 36, 60-286 Pozna{\'n}, Poland \\
$^{17}$School of Physical Sciences, The Open University, Milton Keynes, MK7 6AA, UK\\
$^{18}$Astrophysics Branch, NASA Ames Research Center, Moffett Field, CA, USA\\
$^{19}$Leiden Observatory, Leiden University, P.O. Box 9513, NL-2300 RA Leiden, The Netherlands
}

\date{Accepted 2020 June 8. Received 2020 June 2; in original form 2020 March 17.}

\pubyear{2020}

\begin{document}
\label{firstpage}
\pagerange{\pageref{firstpage}--\pageref{lastpage}}
\maketitle

\begin{abstract}
Using the Eight MIxer Receiver (EMIR) instrument on the Institut de RadioAstronomie Millim\'etrique (IRAM) 30~m telescope, we conducted a spectroscopic redshift search of seven $z_{\rm phot}$ $\sim$ 4 sub-millimetre bright galaxies selected from the \textit{Herschel} Bright Sources (HerBS) sample with fluxes at 500~$\mu$m greater than 80~mJy. For four sources, we obtained spectroscopic redshifts between $3.4<z<4.1$ through the detection of multiple CO-spectral lines with J~$\geq3$. Later, we detected low-J transitions for two of these sources with the GBT including the CO(1-0) transition. For the remaining three sources, more data are needed to determine the spectroscopic redshift unambiguously. The measured CO luminosities and line widths suggest that all these sources are gravitationally lensed. These observations demonstrate that the 2~mm window is indispensable to confirm robust spectroscopic redshifts for $z < 4$ sources. Finally, we present an efficient graphical method to correctly identify spectroscopic redshifts.
\end{abstract}



\section{Introduction}
Initial far-infrared/submillimetre (sub-mm) observations were painfully slow, detecting only a sub-mm source every couple of hours, even when assisted by gravitational lensing \citep{sma97}. The \textit{Herschel Space Observatory} \citep{pil10} changed the detection speed significantly, and increased the number of known sub-millimetre galaxies (SMGs) from hundreds to hundreds of thousands. The H-ATLAS survey (\textit{Herschel} Astrophysical Terahertz Large Area Survey - \citealt{eal10}; \citealt{val16}; \citealt{mad18}) is one of the largest legacies of the \textit{Herschel} mission. This survey observed a total of 616.4 square degrees over five fields in five \textit{Herschel} bands. The large-area surveys done with \textit{Herschel} allow us to select sources that are among the brightest in the sky. After the removal of easily-identifiable interlopers, such as local galaxies and blazars, a large percentage of these sources turn out to be lensed ULIRGs (Ultra-Luminous Infrared Galaxies, $10^{12}$ L$_{\odot}$ < L$_{\rm{FIR}}$ < $10^{13}$ $L_{\odot}$) at high redshift  \citep{neg07, neg10, gon12, gon19}. The few sources that turn out to be only weakly-lensed, in fact, are even more intrinsically-luminous HyLIRGs (Hyper-Luminous Infrared Galaxy, $L_{\rm{FIR}}$ > $10^{13}$ L$_{\odot}$) (e.g. \citealt{ivi13}).

Understanding the nature and physical properties of these sources requires an accurate redshift. Photometric redshifts are only indicative, and are influenced by the internal properties of the galaxy (e.g. dust temperature and dust emissivity index, $\beta$), worsened significantly by the degeneracy between these two parameters (e.g. \citealt{bla99,smi12}). Instead, we require multiple spectral lines to accurately determine the redshift. These then empower us to determine clustering properties of the individual sources, and to follow-up specific spectral lines which are able to characterise the properties of the galaxies, tracing phenomena such as outflows (e.g. \citealt{fal17,spi18}) and star formation (e.g. \citealt{loo11}).

Due to the drawbacks of optical and near-infrared redshift searches (i.e. redshift bias and positional uncertainty; \citealt{dan02,dan04,dan08}), there was a strong push towards the development of spectroscopic instruments in the sub-mm with the relative wide bandwidth necessary for finding spectral lines. Early sub-mm/mm spectroscopic instruments had relative bandwidths ($\delta\lambda$/$\lambda$) of less than 1\%, but the development of Z-Spec \citep{nay03}, the Redshift Search Receiver (RSR;  \citealt{eri07}), Zpectrometer \citep{har07} and EMIR \citep{car12} on single-dish telescopes, and the  {\color{referee}WideX correlator on the Plateau de Bure interferometer (PdBI, later upgraded to the PolyFIX correlator on the NOrthern Extended Millimeter Array; NOEMA, e.g. \citealt{ner20})}, and bands 3 and 4 on the Atacama Large Millimetre/submillimetre Array (ALMA; \citealt{ker08}), pushed the relative bandwidths up to 10\% and beyond. 
This increase in relative bandwidth translates directly to an increase in redshift coverage, and thus redshift searches with sub-mm instruments became viable. Moreover, {\color{referee}far-infrared spectral lines are visible out to high redshift, and the observations are not influenced by the positional uncertainty that plagued earlier redshift searches.}
To summarize, the advent of large bandwidth receivers in the (sub-)millimeter in the past decade finally enabled the start of efficient redshift searches for highly dust-obscured galaxies. 

In the past years several groups have conducted redshift searches \citep[e.g.,][]{wei09,wei13,har10,swi10,vie13,rie13,rie17,can15,zav15,har16,fud17,ner20}, mainly of galaxies selected from large-area surveys observed with {\it Herschel}, {\it Planck} and the South Pole Telescope (SPT). The specific wavelength at which these sources were selected influences the redshift distribution of the sample. In the sub-mm wavelength regime, the cosmological dimming is compensated for by the steep Rayleigh-Jeans part of the spectrum. This, in turn, causes the sub-mm flux density of a source to be only weakly dependent on redshift, although the extent of this so-called \textit{negative K-correction} wavelength-dependent effect depends on the exact wavelength \citep{fra91,bla93}.
In practice, the long-wavelength selection of the SPT ($\sim$1.4\,mm) is nearly independent of redshift, and is able to probe out to redshift $z \sim 6-7$ \citep{mar18}, while the 500\,$\mu$m \textit{Herschel} SPIRE instrument is more sensitive towards sources at lower redshifts around $z \sim 1$ to 5. This allows us to select sources with specific redshift distributions, in order to study the dust-obscured star formation in different eras of cosmic evolution. The peak in cosmic star formation density occurred around redshift 2.5 \citep{mad14}, the \textit{cosmic noon}, while the first dust-obscured objects appeared somewhere beyond redshift 4, the \textit{cosmic dawn}. Neither region, however, has been studied with sufficient detail to trace the evolution of dust-obscured star formation out to high redshift \citep{cas18}, similar to studies of the unobscured star formation using rest-frame optical and UV studies (e.g. \citealt{bou15,fin15}).

This paper reports a redshift search of seven \textit{Herschel}-detected galaxies with the instrument EMIR on the IRAM 30~m telescope. 
The major aim of our work is to identify the spectroscopic redshifts of dusty star-forming {\color{referee}{\it Herschel}-selected} galaxies around $z \sim 4$, between the \textit{cosmic dawn} and \textit{cosmic noon}, using the CO emission lines. These galaxies probe the high-redshift range of \textit{Herschel}-selected sources, as part of a larger effort to identify the spectroscopic redshifts of bright \textit{Herschel} galaxies \citep[e.g.][]{ner20}. 

The paper is structured as follows. In Section \ref{sec:sources} we describe our sample, our observations and the results in Section \ref{sec:observations}. We describe our redshift search method in Section \ref{sec:specZs}, where we discuss the sources individually, and provide an overview of future redshift searches. In Section \ref{sec:discussion}, we discuss the physical properties of our galaxies, following from the observations in this paper. Finally, we provide our conclusions in Section \ref{sec:conclusions}. Throughout this paper, we assume a flat $\Lambda$-CDM model with the best-fit parameters derived from the results from the Planck Observatory \citep{pla16}, which are $\Omega_m$ = 0.307 and $h = 0.693$.

\section{Source selection}
\label{sec:sources}
Our sources are selected from the \textit{Herschel} Bright Sources sample (HerBS; \citealt{bak18,bak20err}), which contains the brightest, high-redshift sources in the 616.4 sqr. deg. H-ATLAS survey \citep{eal10}, in total 209 sources. The H-ATLAS survey used the PACS \citep{pog10} and SPIRE \citep{gri10} instruments on the \textit{Herschel Space Observatory} to observe the North and South Galactic Pole Fields and three equatorial fields to a 1$\sigma$ sensitivity of 5.2 mJy at 250 $\mu$m to 6.8 mJy at 500 $\mu$m\footnote{These sensitivities are derived from the background-subtracted, matched-filtered maps without accounting for confusion noise} \citep{eal10,val16}.  The sources are selected with a photometric redshift, $z_{\rm{phot}}$, greater than 2 and a 500 $\mu$m flux density, S$_{\rm{500\mu m}}$, greater than 80 mJy. Blazar contaminants were removed with radio catalogues and verified with 850 $\mu$m SCUBA-2 observations \citep{bak18}. The far-infrared photometric redshift of each source was derived via fitting of a two-temperature modified blackbody (MBB) SED template --- based on $\sim$40 lensed H-ATLAS sources with spectroscopic redshifts \citep{pea13} --- to the 250, 350 and 500 $\mu$m flux, see for details \citet{bak18}.  

\begin{figure}
\includegraphics[width=\linewidth]{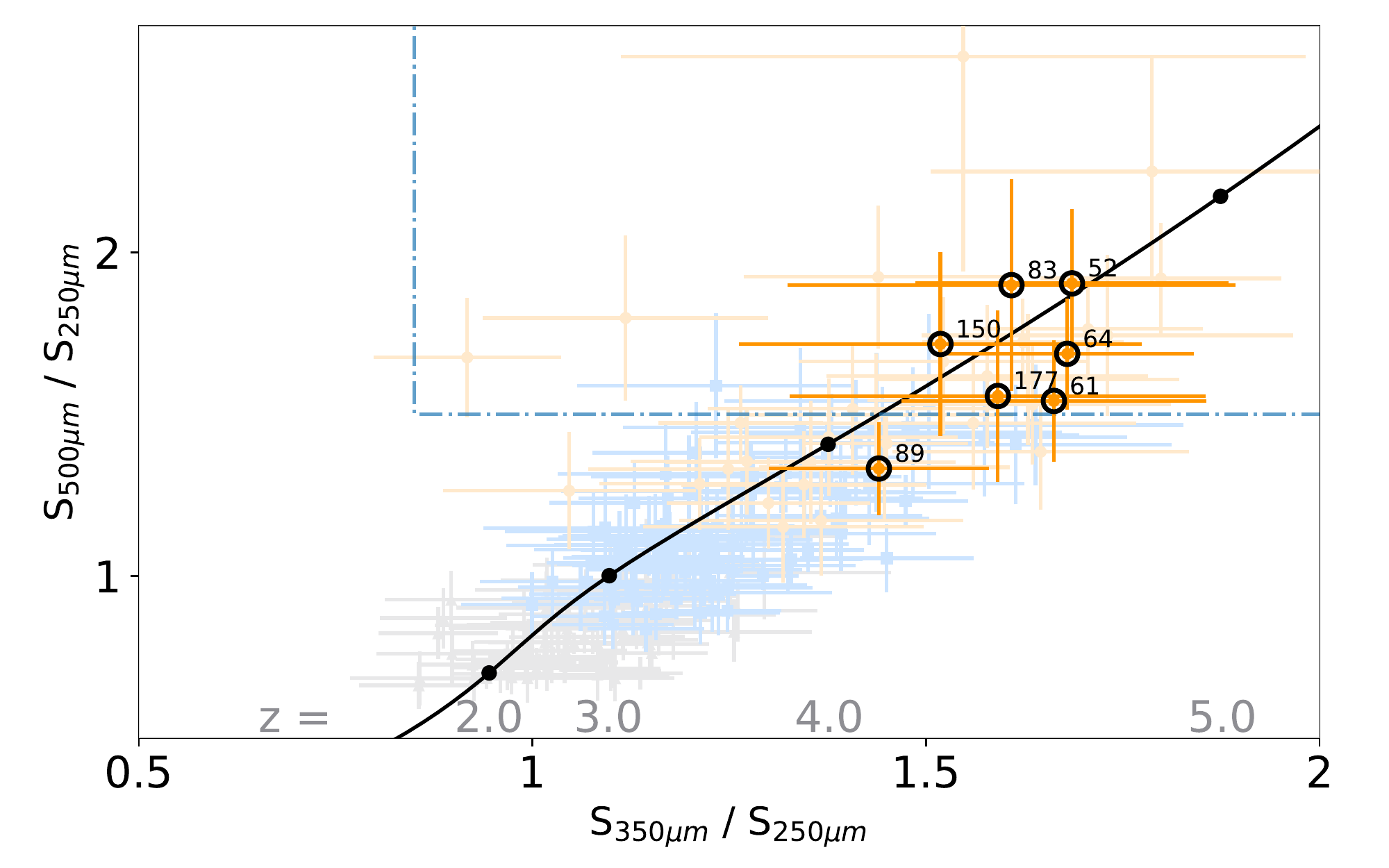}
\caption{The S$_{\rm 500~\mu m}$/S$_{\rm 250~\mu m}$ vs. S$_{\rm 350~\mu m}$/S$_{\rm 250~\mu m}$ colour-colour plot of our selected sources (open, orange dots) from {\it Herschel}/SPIRE observations. The shaded points represent the total HerBS sample (grey, z$_{\rm phot}$ $<$ 2.5; blue, 2.5 $<$ z$_{\rm phot}$ $<$ 3.0;  orange, z$_{\rm phot}$ $>$ 3.0). The solid black line indicates the redshift evolution of the spectrum derived from 24 HerBS sources with spectroscopic redshifts \citep{bak18}, and the dash-dotted line indicates the selection criterion for ultra-red sources \citep{ivi16}.}
\label{fig:sourceselection}
\end{figure}

Therefore, for the follow-up with IRAM 30m telescope, we selected seven sources with photometric redshifts around $z \sim 4$ from the NGP and GAMA fields, both accessible from the IRAM 30~m telescope, and having clear (S$_{\rm 850\mu m}$ $>$ 40 mJy) detections with SCUBA-2 at 850~$\mu$m (see Table~\ref{tab:photprop} for their properties and see Figure~\ref{fig:sourceselection} for the position of the sources on a colour-colour plot). Note that the 850$\mu$m SCUBA-2 fluxes and photometric redshifts are from the updated SCUBA-2 catalogue in \cite{bak20err}, where the SCUBA-2 fluxes were recalculated and revised from \cite{bak18}.

Three of these sources are also found in the catalogue of potentially-lensed sources in \cite{neg17}. Using SDSS follow-up, they find it likely that HerBS-52 is gravitationally lensed, while their analysis remains inconclusive on sources HerBS-61 and HerBS-64. 
A near-infrared survey done with the VISTA-telescope (VISTA Kilo-Degree Infrared Galaxy Survey --- VIKING; \citealt{edg13}) covers four of these sources. \cite{bak20} has shown that HerBS-61, -83, -150 and -177 have a nearby bright VIKING galaxy, and derived the probability that these VIKING galaxies are associated with the background \textit{Herschel} source. Each of these near-infrared galaxies have a photometric redshift derived from its five-band near-infrared colours. HerBS-61 and HerBS-177 have a high --- 97\% and 99\% --- chance of the VIKING galaxy being an associated source. The photometric redshifts of these two VIKING galaxies are $z \approx 0.7$. HerBS-83 has a nearby VIKING galaxy with an association probability of 57\%, and its VIKING redshift is $z \approx 1.1$. HerBS-150 has a nearby VIKING galaxy with an association probability of 69\%, and its VIKING redshift is $\approx 0.7$. These associations increase our suspicion that these sources are gravitationally lensed.

\begin{table*}
\caption{IRAM 30~m sample}
\label{tab:photprop}
\resizebox{\textwidth}{!}{\begin{tabular}{lcccccccccc}  
\hline
\hline
Source & H-ATLAS name       & RA & DEC                  & S$_{\rm{250\mu m}}$ & S$_{\rm{350\mu m}}$ & S$_{\rm{500\mu m}}$ & S$_{\rm{850\mu m}}$ & z$_{\rm{phot}}$ & log $\mu$L$_{\rm{FIR}}$ & $\mu$ SFR \\
 &                          & [hms] & [dms]             & [mJy]         & [mJy] & [mJy] & [mJy] &  & [log L$_{\odot}$] & [10$^3 \times{}$\,M$_{\odot}$/yr]\\ \hline
HerBS-52 & J125125.8+254930 & 12:51:25.8 & 25:49:30     & 57.4 $\pm$ 5.8 & 96.8 $\pm$ 5.9 & 109.4 $\pm$ 7.2 & 46.5 $\pm$ 9.1    & 3.7 $\pm$ 0.5 & 13.59 & 5.8 $\pm$ 0.5 \\
HerBS-61 & J120127.6-014043 & 12:01:27.6 & -01:40:43    & 67.4 $\pm$ 6.5 & 112.1 $\pm$ 7.4 & 103.9 $\pm$ 7.7 & 45.1 $\pm$ 7.8   & 3.2 $\pm$ 0.4 & 13.51 & 4.8 $\pm$ 0.5 \\
HerBS-64 & J130118.0+253708 & 13:01:18.0 & 25:37:08     & 60.2 $\pm$ 4.8 & 101.1 $\pm$ 5.3 & 101.5 $\pm$ 6.4 & 50.7 $\pm$ 6.7   & 3.9 $\pm$ 0.5 & 13.65 & 6.6 $\pm$ 0.5 \\
HerBS-83 & J121812.8+011841 & 12:18:12.8 & 01:18:41     & 49.5 $\pm$ 7.2 & 79.7 $\pm$ 8.1 & 94.1 $\pm$ 8.8 & 48.2 $\pm$ 8.4     & 3.9 $\pm$ 0.5 & 13.57 & 5.5 $\pm$ 0.5 \\
HerBS-89 & J131611.5+281219 & 13:16:11.5 & 28:12:19     & 71.8 $\pm$ 5.7 & 103.4 $\pm$ 5.7 & 95.7 $\pm$ 7.0 & 48.7 $\pm$ 7.0    & 3.5 $\pm$ 0.5 & 13.59 & 5.8 $\pm$ 0.5 \\
HerBS-150 & J122459.1-005647 & 12:24:59.1 & -00:56:47   & 53.6 $\pm$ 7.2 & 81.3 $\pm$ 8.3 & 92.0 $\pm$ 8.9 & 41.3 $\pm$ 7.4     & 3.4 $\pm$ 0.4 & 13.45 & 4.2 $\pm$ 0.4 \\
HerBS-177 & J115433.6+005042 & 11:54:33.6 & 00:50:42    & 53.9 $\pm$ 7.4 & 85.8 $\pm$ 8.1 & 83.9 $\pm$ 8.6 & 48.0 $\pm$ 7.0     & 3.9 $\pm$ 0.5 & 13.59 & 5.8 $\pm$ 0.5 \\ \hline
\end{tabular}}
\raggedright \justify \vspace{-0.2cm}
\textbf{Notes:} Col.~(1) : HerBS-ID. Col.~(2) : Official H-ATLAS name \citep{val16,bou16}. Col.~(3) and (4) : SPIRE 250~$\mu$m positions. Col.~(5), (6) and (7) : SPIRE fluxes at 250, 350 and 500~$\mu$m respectively. Col.~(8) : SCUBA-2 fluxes at 850~$\mu$m. Col.~(9): Photometric redshift, assuming an uncertainty of d$z/(1+z)$ = 13\% from \cite{bak18}. Col~(10): Far-infrared luminosity derived from \textit{best-fit} template of \citet{bak18}. Col.~(11) : star formation rate based on \citet{ken13} assuming the far-infrared luminosity from Col~(10). 
\end{table*}

\section{Observations and Data Reduction}
\label{sec:observations}
\subsection{IRAM EMIR}
Using the multi-band heterodyne receiver EMIR, we started our redshift search at IRAM in the 3~mm window (E0: $73-117$~GHz). This window is guaranteed to contain at least one CO line for the redshift ranges 1.2 $<$ z $<$ 2.15 and z $>$ 2.3, where our sources are expected to lie. After the first line detection, we retune mainly to the 2~mm window (E1: $125-184$~GHz) to confirm the spectroscopic redshift via the detection of a second CO line. As backend we used the fast Fourier Transform Spectrometer (FTS200) with a 200~kHz resolution. Observations were carried out mainly in visitor mode, as real-time line confirmations and/or adjustments of the observing strategy were essential. The telescope was operated in position-switching mode, in order to subtract atmospheric variations. Our IRAM observations were carried out in four separate programs, from 2015 until 2017; 080-15 (PI: Helmut Dannerbauer), 079-16, 195-16 and 087-18 (PI: Tom Bakx). These observations took place with good to acceptable weather conditions, for details see Table \ref{tab:obsprop}.

\begin{table*}
\caption{IRAM 30~m observations}
\label{tab:obsprop}
\begin{tabular}{lcccc}
\hline\hline
Source    & Dates & Time per setup [h] & tau$_{\rm{eff}}$ \\ \hline
HerBS-52  & 28+29/07/15 03/08/15 & 13.4 (E0: 9.7; E1: 3.7) & 0.03 - 0.6 \\
HerBS-61  & 06+27/07/16 20+21/10/16 10/04/17 & 9.2 (E0: 4.0; E1: 5.2) & 0.05 - 0.8 \\
HerBS-64  & 20+21/05/17 29:31/07/15 01/08/15 & 12.8 (E0: 10; E1: 2.8) & 0.02 - 0.6 \\
HerBS-83  & 26/07/16 20/10/16 27/11/16 & 15.5 (E0: 14.85; E1: 0.6) & 0.02 - 0.4 \\
      & 06+09+13/04/17 16:22+23/05/17 \\
HerBS-89  & 20/05/17 & 1.6 (E0: 1.6) & 0.05 - 0.15 \\
HerBS-150   & 06/07/16 20+21/10/16 12+13/04/17 21/05/17 & 5.0 (E0: 4.0, E0+E1: 1.0) & 0.05 - 0.58 \\
HerBS-177   & 06+11/07/16 20+21/10/16 28/11/16 06+12/04/17 & 9.8 (E0: 7.3; E1: 2.5) & 0.03 - 0.6 \\ \hline
\end{tabular}
\raggedright \justify \vspace{-0.2cm}
\textbf{Notes:} Col- ~(1): HerBS-ID.  Col.~(2): Dates of observation. Col.~(3): Observing time per source and band.  Col-~(4): Tau corrected for elevation.
\end{table*}

We reduced the IRAM data using the GILDAS package CLASS and Python, using the modules numpy \citep{vir19}, astropy \citep{ast13,ast18} and lmfit \citep{new19}. Initially, we used CLASS to check the data during the observations. Then we used Python to remove the baseline of individual scans, and to remove bad data. After that, we combined the data and redid the baseline removal while masking out the positions of the spectral lines in order to get accurate flux estimates of the spectral lines. Each six minute scan consists of eight 4 GHz subbands, split by horizontal and vertical polarisation. In total, each scan covered 16 GHz --- 8 and 8 GHz separated by 8 GHz --- in both horizontal and vertical polarisation in one scan.  After the baseline-subtraction, we removed bad data by visual inspection of each subband (up to 10\% of the data was removed). Typical examples of when a subscan was considered \textit{bad data} include strong sinusoidal interference, a strong peak, or a baseline that did not appear linear. All good data were then combined into a spectrum with a frequency resolution of $\sim$70~km~s$^{-1}$, weighted by both the amount of data points which mapped onto this frequency binning, and by the noise level of the subscan:

\begin{equation}
\rm S_j = \frac{ \sum^{k}{\frac{\sum^{i \in j}{N_{i,k}}}{\rm{VAR} ({N_k})} }} {\sum^k{n_{i \in j} / {VAR} (N_k) }}.
\end{equation}
Here, $\rm S_j$ is the lower-resolution spectrum for a source, where j is the frequency index. The subband data, $N_{i,k}$, has $i$ data points, for all $k$ subbands available for each source. VAR$(N_k)$ is the variance of the subband data (equal to the square of the standard deviation), and $n_{\rm i \in j}$ is the number of data points $i$ that map onto each lower-resolution frequency bin $j$.

The binned spectra increase the per-bin signal-to-noise, and facilitate the identification of spectral lines. Once we found the location of a spectral line in a spectrum, we return to the raw data in order to remove the baseline while masking out the position of the spectral line. We convert from temperature to flux using point source conversion factors at 3~mm (5.9~Jy/K) and 2~mm (6.4~Jy/K).  A typical absolute flux calibration uncertainty of 10\% is also taken into account.

\subsection{GBT VEGAS}
Observations of the lower-J CO transitions of HerBS-52 and HerBS-64, {\color{referee} for which robust redshifts were measured with EMIR},  were carried out using the Robert C. Byrd Green Bank Telescope (GBT).  Five observing sessions were conducted in 2016 October through December (GBT program: 16B210, PI: H. Dannerbauer), and two observing sessions were carried out in 2018 March (GBT DDT program: 18A459, PI: T. Bakx). For HerBS-52, we observed CO(1-0) in K-band and CO(3-2) in W-band (Table~3), while for HerBS-64 CO(1-0), CO(2-1), and CO(3-2) were observed in K, Q, and W-band respectively. All observations used the Versatile GBT Astronomical Spectrometer (VEGAS) with a bandwidth of 1500\,MHz and a raw spectral resolution of 1.465\,MHz, which provided sufficient velocity coverage and velocity resolution for all bands. For the K- and W bands, the NOD observing mode was adopted.  This mode alternates observations between two beams by moving the telescope. For the Q-band, we used the SubBeamNod observing mode which moves the sub-reflector to alternate observations between two beams. SubBeamNod observations allow for faster position switched observations and yield better baselines for receivers whose beam separations are small (e.g., the Ka-band, Q-band, and Argus instruments on the GBT).  For K- and W-bands, NOD observations yield better baselines. A nearby bright continuum source was used to correct the pointing and focus of the telescope every 30 to 60 minutes, depending on the observing frequency and conditions. For observations with Q and W-band, the surface thermal corrections for the telescope were made using the AutoOOF observations of 3C273. This method adjusts the surface actuators of the telescope for the current conditions to improve the aperture efficiency at high frequency. The pointing of the telescope as well as the methods for correcting the surface work best in night time under stable thermal conditions. The CO(3-2) observations for HerBS-52 were taken in the afternoon under very clear, sunny skies which greatly degraded the quality of these data. During the afternoon with clear skies, the changing of the surface and the structure yield large effective telescope losses at W-band.

The GBT spectral-line data were reduced using GBTIDL. Each individual NOD and SubBeamNod scan was visually inspected for the two polarisations and two beams. Low-order polynomial baselines were removed and scans showing large residual baseline features were removed before data co-addition. The data were corrected for atmospheric losses. This correction was particularly large for the CO(3-2) data of HerBS-64 since the observed frequency of 68.6 GHz is within the wing of the strong atmospheric O$_2$ band.  The data were corrected for drifts in pointing by using the measured pointing offsets and assuming a Gaussian beam. The absolute flux density scales for the K-band and Q-band observations were derived from observations of 3C286 based on the VLA calibration results of \citet{per13}. For W-band, we used observations of 3C273 and the known flux density as a function of time provided by the ALMA Calibrator Source Catalog online database. Factoring in estimates of all errors, including the uncertainty on the calibration scales, measurement errors, uncertainty for the atmosphere correction, and the uncertainty associated with the pointing and focus drifts, we estimated flux uncertainties errors of 15\% at K-band, 20\% at Q-band, 40\% at W-band for HerBS-64, and 60\% at W-band for HerBS-52.  The calibration at W-band was particularly challenging for HerBS-52 due to the sunny afternoon conditions which significantly impacted the ability to derive accurate telescope corrections.

\subsection{Spectral line measurements}
\begin{figure*}
\centering
\includegraphics[width=0.75\textheight]{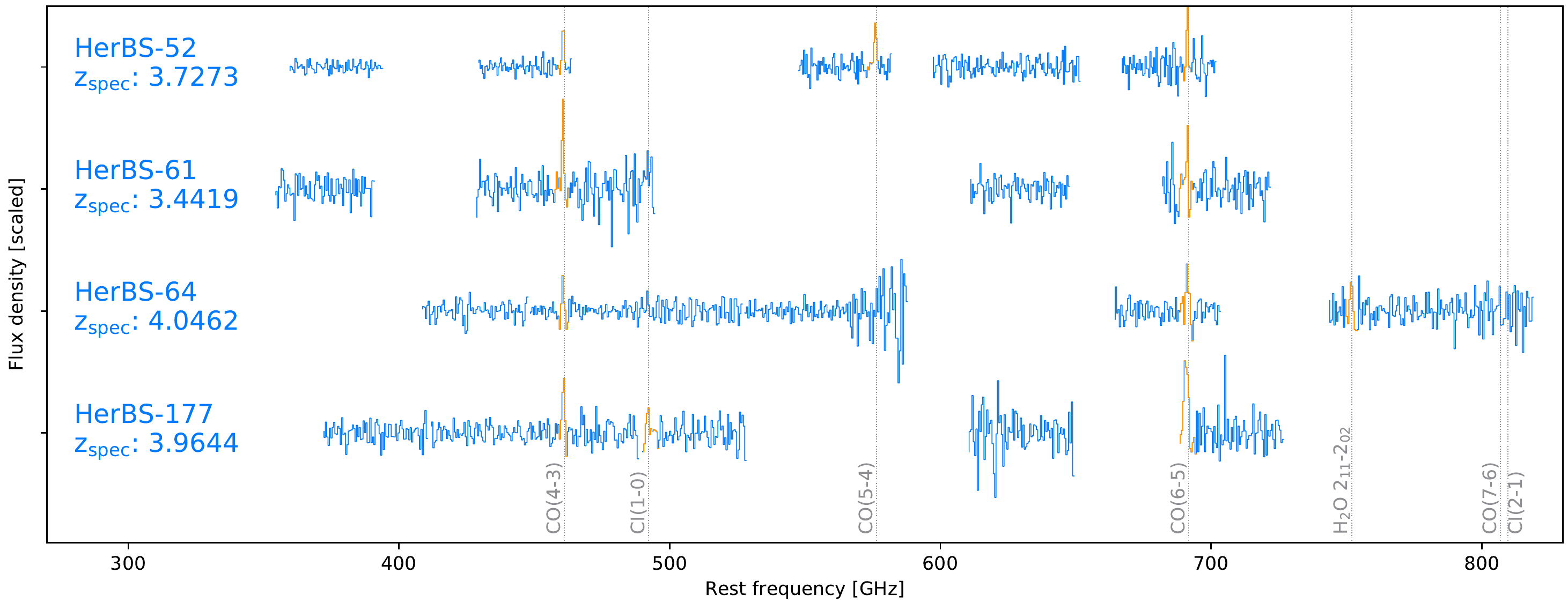}
\caption{Rest-frame sub-mm spectra of the four sources with confirmed spectroscopic redshifts obtained with the EMIR instrument at the IRAM 30~m telescope. The orange lines mark the detected spectral lines of each galaxy.}
\label{fig:combinedPlots}
\end{figure*}
\label{sec:results}
{\color{referee2}We provide all spectra from the IRAM 30m and GBT observations in the Appendix Figures~\ref{fig:HerBS52} to \ref{fig:HerBS177}, where we show the spectra at 70~km/s in individual polarisations (\textit{grey}) and combined polarisations (\textit{blue}).
We initially identified the lines by eye, ensuring the line is detected in both polarisations ({\it grey lines} in Fig. \ref{fig:HerBS52} to \ref{fig:HerBS177}).}
We extract the line properties from the EMIR spectrum binned at a velocity resolution of 70 km/s. In order to estimate the uncertainty on our spectral line parameters, we create 1000 realisations of the spectrum to which we have added artificial noise, based on the standard deviation of the spectrum away from the spectral line. We fit a Gaussian profile to each of these 1000 realisations, which gives us one thousand estimates for each line parameter: frequency, integrated flux and velocity width. We fit the distribution of the thousand data points for each line parameter by a Gaussian, where the central position of the Gaussian is our best estimate for the parameter, and the standard-deviation of this Gaussian function is an accurate measure for the uncertainty. {\color{referee2}This allowed us to extract the line properties of detections down to a signal-to-noise ratio of $\gtrsim3\sigma$ from the velocity-integrated flux density.}

We present the results of the observations in Table \ref{tab:specdef}. For four sources, we obtained a secure spectroscopic redshift based upon multi-line detections, which always required observations in both the 3 and 2~mm windows. Almost all detected lines are from CO, and in addition we detect for one source the [C\,{\sc i}](1-0) line and for one source a water line (H$_{\rm 2}$O 2$_{\rm 11}$-2$_{\rm 02}$). For the J $>3$ CO observations at the IRAM 30m, we derive velocity-integrated fluxes I$_{\rm CO}$ from 1.5 to 13 Jy km~s$^{-1}$, while our lower-J transitions (J $\leq$ 3), observed with the GBT range between 0.1 and 0.8 Jy km s$^{-1}$. {\color{referee}We do not correct for CMB effects (e.g. observation contrast and enhanced CO excitation), since this requires extensive modelling of the ISM conditions, and exceeds the scope of this paper \citep{dac13,zha16}.}
We show the rest-frame spectra of the sources with spectroscopic redshifts in Figure~\ref{fig:combinedPlots}. The velocity-width FWHM of our sources ranges between 250 and 750 km/s. For each galaxy with a confirmed spectroscopic redshift, we detect the CO(4-3) transition. {\color{referee2}For three sources of our sample, we were not able to identify a robust spectroscopic redshift, given the spectral lines detected for each source. Instead, we list their spectral line properties in Table \ref{tab:nonzspec}.}

\begin{table*}
\caption{Confirmed spectral lines}
\label{tab:specdef}
\resizebox{\linewidth}{!}{\begin{tabular}{lcccccc} 
\hline\hline
Source    & Redshift & Transition & Frequency [GHz] & I$_{\rm CO}$ [Jy km/s]  & FWHM [km/s] & S/N\\ 
\hline
HerBS-52  & 3.4419 $\pm$ 0.0006 & CO(1-0) & 25.951 $\pm$ 0.004 & 0.39 $\pm$ 0.06 & 519 $\pm$ 85 & 6.5 \\
          & & CO(3-2) & 77.854 & $<$ 3.52 & & \\
          & 3.4428 $\pm$ 0.0005 & CO(4-3) & 103.783 $\pm$ 0.012 & 6.07 $\pm$ 0.70 & 507 $\pm$ 52 & 8.7 \\
          & 3.4423 $\pm$ 0.0006 & CO(5-4) & 129.744 $\pm$ 0.017 & 6.52 $\pm$ 0.95 & 515 $\pm$ 72 & 6.9 \\
          & 3.4409 $\pm$ 0.0005 & CO(6-5) & 155.742 $\pm$ 0.016 & 6.22 $\pm$ 1.52 & 229 $\pm$ 33 & 4.1 \\
\hline
HerBS-61  & 3.7275 $\pm$ 0.0005 & CO(4-3) & 97.532 $\pm$ 0.011 & 5.66 $\pm$ 0.87 & 456 $\pm$ 88 & 6.5 \\
          & & [C\,{\sc i}](1-0) & 104.086 $\pm$ 0.066 & $<$ 3.63 &  \\
          & 3.7271 $\pm$ 0.0006 & CO(6-5) & 146.312 $\pm$ 0.018 & 2.25 $\pm$ 0.62  & 247 $\pm$ 41 & 3.6 \\
\hline
HerBS-64  & 4.0484 $\pm$ 0.0009 & CO(1-0) & 22.833 $\pm$ 0.004 & 0.18 $\pm$ 0.05 & 310 $\pm$ 81 & 3.6 \\
          & 4.0462 $\pm$ 0.0006 & CO(2-1) & 45.686 $\pm$ 0.005 & 0.51 $\pm$ 0.12 & 455 $\pm$ 79 & 4.3 \\
          & & CO(3-2) & 68.630 $\pm$ 0.017 & $<$ 0.49 & \\
          & 4.0473 $\pm$ 0.0007 & CO(4-3) & 91.352 $\pm$ 0.013 & 2.06 $\pm$ 0.53 & 364 $\pm$ 99 & 3.9 \\
          & 4.0471 $\pm$ 0.0005 & CO(6-5) & 137.035 $\pm$ 0.015 & 2.40 $\pm$ 0.58 & 331 $\pm$ 123 & 4.1 \\
          & 4.0431 $\pm$ 0.0009 & H$_{\rm 2}$O 2$_{\rm 11}$-2$_{\rm 02}$ & 149.122 $\pm$ 0.027 & 1.60 $\pm$ 0.54 & 345 $\pm$ 73 & 3.0 \\
          & & CO(7-6) & 159.902 & $<$ 1.69 & & \\
          & & [C\,{\sc i}](2-1) & 160.457 & $<$ 1.69 & & \\
\hline
HerBS-177 & 3.9633 $\pm$ 0.0006 & CO(4-3) & 92.898 $\pm$ 0.011 & 4.35 $\pm$ 1.03 & 490 $\pm$ 204 & 4.2 \\
          & 3.9616 $\pm$ 0.0009 & [C\,{\sc i}](1-0) & 99.187 $\pm$ 0.018 & 2.00 $\pm$ 0.60 & 308 $\pm$ 102 & 3.3 \\
          & 3.9673 $\pm$ 0.0007 & CO(6-5) & 139.236 $\pm$ 0.019 & 7.05 $\pm$ 0.70 & 677 $\pm$ 62 & 10 \\ 
\hline
\end{tabular}}
\raggedright \justify \vspace{-0.2cm}
\textbf{Notes:} Col.~(1): HerBS-ID. Col.~(2): spectroscopic redshift.  Col.~(3):  spectral line. Col.~(4): observed frequency. Col.~(5): integrated flux of spectral line, for unobserved lines we assume a line width of 500 km/s and a 3 $\sigma$ upper limit. Col.~(6): line width expressed in full-Width at half-maximum (FWHM). {\color{referee2}Col.~(7): signal-to-noise ratio of the spectral line.}
\end{table*}

\begin{table}
\caption{ {\color{referee}Unconfirmed spectral lines}}
\label{tab:nonzspec}
\resizebox{\linewidth}{!}{\begin{tabular}{lcccc} 
\hline\hline
Source    & Frequency [GHz] & I$_{\rm CO}$ [Jy km/s]  & FWHM [km/s] & S/N \\ 
\hline
HerBS-83  & 89.594 $\pm$ 0.018 & 2.98 $\pm$ 0.49 & 636 $\pm$ 82 & 6.1  \\
          & 93.235 $\pm$ 0.070 & 1.84 $\pm$ 0.69 & 255 $\pm$ 96 & 2.7 \\
          & 101.759 $\pm$ 0.022 & 2.57 $\pm$ 0.54 & 539 $\pm$ 93 & 4.8 \\ \hline
HerBS-150 & 83.466 $\pm$ 0.024 & 13.1 $\pm$ 3.63 & 635 $\pm$ 117 & 3.6 \\
          & 94.089 $\pm$ 0.288 & 2.06 $\pm$ 0.78 & 580 $\pm$ 111 & 2.6 \\
\hline
\end{tabular}}
\raggedright \justify \vspace{-0.2cm}
\textbf{Notes:} Col.~(1): HerBS-ID. Col.~(2): observed frequency. Col.~(3): integrated flux of spectral line, for unobserved lines we assume a line width of 500 km/s and a 3 $\sigma$ upper limit. Col.~(4): line width expressed in full-Width at half-maximum (FWHM). {\color{referee2}Col.~(5): signal-to-noise ratio of the spectral line.}
\end{table}

\section{Spectroscopic redshifts} 
\begin{figure*}
\centering
\includegraphics[width=\linewidth]{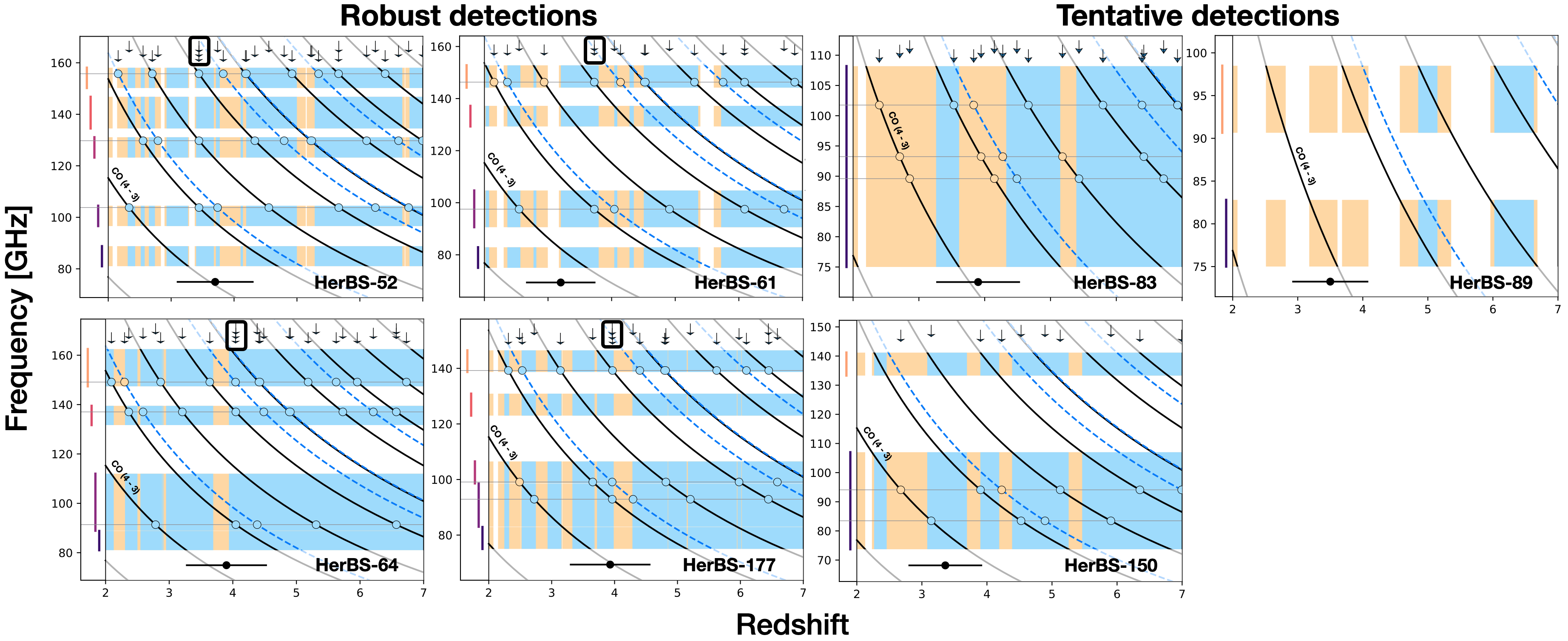}
\caption{The observed frequency against the covered redshift for all sources, separated between robustly-detected redshifts (\textit{left-hand side}) and ambiguous redshifts (\textit{right-hand side}). For each potentially-observed spectral line (CO: \textit{black line}, H$_2$O and [C\,{\sc i}]: \textit{dashed blue line}), we show their observed frequency as a function of redshift. On the left-hand-side, we show the observed bandwidth, and indicate at which redshifts we would have expected to detect 1 CO-line (\textit{orange fill}) or multiple CO-lines (\textit{blue fill}). We indicate the frequencies of the observed spectral lines with horizontal lines, and indicate the corresponding potential redshifts for each line with an arrow in the top of the graph, {\color{referee} with the photometric redshift estimate at the bottom of each graph}. For the robust detections, we note that these arrows line up at the robust redshift identification, encircled with a \textit{black line}. If the redshift is not robustly-identified, the redshift possibilities within the \textit{blue fill}, e.g. expected multiple CO-line detection, can be excluded, as this would have resulted in multiple line detections. This graphical method removes the need for unwieldy and error-prone comparisons of spectral lines, and graphically shows the frequencies that need to be probed in future missions. }
\label{fig:search}
\end{figure*}
\label{sec:specZs}
We discuss an efficient graphical analysis to identify the spectroscopic redshift solutions of our sources in Section \ref{sec:GraphAnalysis}. In Section \ref{sec:indvSources}, we implement this method on the sources with a robust redshift solution and discuss the sources individually. In Section \ref{sec:nozspec}, we discuss the sources without a robust spectroscopic redshift identification, and provide potential explanations for their non-identification. We conclude this section with a brief discussion on future spectroscopic redshift searches in Section \ref{sec:CurrentFutureZspec}.

\subsection{Graphical analysis for unambiguous spectroscopic redshift identification}
\label{sec:GraphAnalysis}
{\color{referee2}We adopt a relatively conservative approach in our identification of our spectroscopic redshifts, demanding a robust spectroscopic redshift via at least two spectral lines.} This requires the cross-comparison of multiple redshift possibilities for each detected spectral line. 
Numerically solving this cross-comparison is a tedious approach, prone to errors and overlooked options. Instead, we use the following graphical method for identifying the potential spectroscopic redshifts for each line. Moreover, the method can be used for sources with only one line detection in order to determine the frequency ranges that need to be probed in future redshift searches.

In Figure~\ref{fig:search}, we show this graphical method for all observed sources as a collage, separated between the sources with unambiguous (\textit{left}) and ambiguous redshift identifications (\textit{right}). In each plot, we show the observed frequency against the covered redshift. We expect to detect CO-lines, and potentially [C\,{\sc i}] and H$_2$O lines, which we graph as \textit{black} and \textit{dashed blue} lines, respectively. Each actually-observed spectral line is indicated with a \textit{horizontal line} at the observed frequency, and where they cross the redshifted spectral lines, we mark the potential spectroscopic redshift with a circle and identify this potential spectroscopic redshift with an arrow in the top of the graph. On the left-hand side, we indicate the observed frequency bands within which we expect to detect all CO-lines. For this range of frequency coverage, we indicate whether the observations would be able to observe one or more spectral lines, indicated with \textit{orange} and \textit{blue fill}, respectively. We do this only for the CO lines, as we cannot be sure we would have detected the [C\,{\sc i}] and H$_2$O emission. The \textit{black point} below indicates the photometric redshift estimate, with its typical uncertainty in $\Delta z / (1+z)$\,$\sim$\, of 13\% \citep{bak18}.

The top arrows line up when multiple lines agree on a robust redshift, identifying the spectroscopic redshift, marked for the sources with unambiguous redshifts with a \textit{black border}. This check is important, because CO-line redshift surveys can suffer from potential multiple redshift solutions, since the CO-ladder increases linearly with each transition J$_{\rm up}$. For example, the observation of J$_{\rm{up}}$ = 2 and 4 CO-line transitions of a z = 2 galaxy (76.7 and 153 GHz, resp.) could also be interpreted as the J$_{\rm{up}}$ = 3 and 6 CO-line transitions of a z $\approx$ 3.5 galaxy. This mis-identification is a non-negligible possiblity given the 13\% errors in sub-mm derived photometric redshifts of around $\Delta z$\,$\sim$\,1 (e.g. \citealt{pea13,ivi16,bak18}). Using Figure~\ref{fig:search}, we can confidently exclude other redshift possibilities for the sources with robust redshift identifications, or identify the need for extra observations to be carried out.

\subsection{Robust redshifts}
\label{sec:indvSources}
\begin{table}
\centering
\caption{Physical properties of sources with robust spectroscopic redshifts.}
\label{tab:specredshifts}
\resizebox{\linewidth}{!}{\begin{tabular}{lccccc}
\hline\hline
Source & z$_{\rm{spec}}$ & $\Delta$z/(1+z) & $\mu$L$_{\rm{FIR}}$ & T$_{\rm dust}$ & $\mu$M$_{\rm dust}$ \\
   & & & log [L$_{\odot}$] & [K] & [10$^9$ M$_{\odot}$] \\
\hline
HerBS-52 & 3.4419 $\pm$ 0.0002  &  -0.058     & 13.53  & 30.7 & 2.7 \\
HerBS-61 & 3.7273 $\pm$ 0.0004  &  0.120     & 13.60  & 34.6 & 1.9 \\
HerBS-64 & 4.0462 $\pm$ 0.0003  &  0.021     & 13.66  & 34.5 & 2.7 \\
HerBS-177 & 3.9644 $\pm$ 0.0004 &  0.007     & 13.58  & 32.7 & 2.9 \\
\hline
\end{tabular}}
\raggedright \justify \vspace{-0.2cm}
\textbf{Notes:} Col.~(1): HerBS-ID. Col.~(2): Spectroscopic redshift calculated from all spectral lines using equation \ref{eq:specz}. Col.(3): Error in photometric redshift estimate, given by (z$_{\rm spec}$ - z$_{\rm phot}$)/(1+z$_{\rm spec}$). Col.~(4): Logarithm of the far-infrared bolometric luminosity, integrated from 8 to 1000 $\mu$m. Col.~(5): Average temperature, assuming a single temperature grey-body with $\beta$ = 2. 
\end{table}
For each source with a consistent spectroscopic redshift identification, we estimate the average spectroscopic redshift, $\bar{z}$, based on a weighted average of the redshifts of each of the lines, $\rm z_i$, combined with the redshift uncertainty in each line, $\rm dz_i$,
\begin{equation}
\rm \bar{z} = \sqrt{\frac{\sum^i{\left( z_i/dz_i \right)^2}}{\sum^i \left( 1/dz_i \right)^2}}.
\label{eq:specz}
\end{equation}
We list spectroscopic redshift per source in Table~\ref{tab:specredshifts}. 

\noindent
\textbf{HerBS-52:} 
We initially detected a spectral line at 103.783~GHz in the E0 band, which placed the galaxy at either $z=2.3, 3.4,$~or~$4.5$. The detections of two lines in the E1 range, at 129.744~GHz and 155.742~GHz, yielded a robust spectroscopic redshift of $z_{\rm CO}=3.4419$. This established the detected lines as CO(4-3), CO(5-4), and CO(6-5).
This is one of two sources in our sample with subsequent GBT follow-up of the CO(1-0), and CO(3-2) transitions. {\color{referee}We detected the CO(1-0), however} the GBT observations of the CO(3-2) line had poor baselines, and hence failed to result in a detection. As a result of this, the flux upper limit we derive for the CO(3-2) line of HerBS-52 is too low, as we will discuss in Section \ref{sec:COSLED}. 

\noindent
\textbf{HerBS-61:}  
We initially detected a spectral line at 97.532~GHz in our first tuning in E0, which placed the galaxy at either $z=2.5, 3.7$~or~$4.9$.
In the subsequent E1 tuning we detected a spectral line at 146.312~GHz, resulting in a spectroscopic redshift of $z_{\rm CO}=3.7273$. This established the detected lines as CO(4-3) and CO(6-5). We are also able to provide an upper-limit on the neighbouring [C\,{\sc i}](1-0) line integrated flux. {\color{referee2}Note that we can reject the redshift solution at $z=6.1$, since there is no emission line at $\sim$81.3~GHz.}

\noindent
\textbf{HerBS-64:} 
We initially detected a spectral line at 91.352~GHz after two tunings in the E0 band, which placed the galaxy at either $z=2.8$~or~$4.0$.
This was followed by two tunings in E1, where we detected a spectral line at 137.035~GHz, which yielded a spectroscopic redshift of $z_{\rm CO}=4.0462$. 
This established the detected spectral lines as CO(4-3) and CO(6-5). We are also able to provide an upper limit for the CO(7-6) and [C\,{\sc i}](2-1) line. 
This is one of two sources in our sample with subsequent GBT follow-up of the CO(1-0), CO(2-1), and CO(3-2) transitions.
The CO(1-0) and CO(2-1) lines are detected clearly with the GBT, however the observations of the CO(3-2) line of HerBS-64 were impacted by high T$_{\rm sys}$ and opacity, since the line is on the wing of the atmospheric O$_2$ line. We failed to detect the CO(3-2) emission, and thus we decided to leave this spectral line out in our further analysis.

\noindent
\textbf{HerBS-177:} 
We initially detected a spectral line at 92.898~GHz after two tunings in the E0 band, with a $\sim$3$\sigma$ detection at 99.187~GHz, which placed the galaxy at $z=3.9$, although $z=2.7$~and~$5.2$ were also possible.
In the subsequent E1 tuning we detected a spectral line at 139.236~GHz, yielding a spectroscopic redshift of $z_{\rm CO}=3.9644$. We excluded a potential solution at z = 4.8 because of the lack of CO-emission at $\sim$79 GHz.
This established the detected spectral lines as CO(4-3), [C\,{\sc i}](1-0), and CO(6-5).

\subsection{HerBS sources without spectroscopic redshift}
\label{sec:nozspec}

\noindent
\textbf{HerBS-83:} 
We observed both tunings in E0 for this source for a significant time (14.7 h in total), revealing several spectral lines. We detected three spectral lines over the entire E0 band, at 89.594, 93.235, and 101.759~GHz. As can be seen in Figure\,\ref{fig:search}, none of these lines agree with a single redshift solution.
In the case that each bandwidth was deep enough to detect all possible spectral lines, we can exclude all \textit{blue} redshift regions in Figure \ref{fig:search}, in which we would have at least detected two spectral lines. This removes the z = 4.3 and 3.4 possibilities.

\noindent
\textbf{HerBS-89:} 
The lack of observing time -- 1.6 hours in a single E0 tuning -- on this source {\color{referee} did not result in a significant line detection.} However, a recent NOEMA study of 13 bright \textit{Herschel}-selected sources by \cite{ner20} revealed the spectroscopic redshift {\color{referee}of HerBS-89} to be $z=2.950$.

\noindent
\textbf{HerBS-150:} 
The first E0 tuning was observed significantly longer than the other tuning in E0 (4h vs. 1h). The deeper tuning revealed a potential spectral line at 83.466~GHz (signal-to-noise (S/N) $\sim$ 3.5), and the shallow observation also detected a spectral line at 94.089~GHz (S/N $\sim$ 3). The frequencies of these spectral lines, however, do not agree with a single redshift. The E1 tuning has not been observed long enough for us to exclude any of the redshift possibilities.

\vspace{0.2cm}
\noindent 
For three of our sources, we could not find a conclusive spectroscopic redshift. For HerBS-89, this is due solely to a lack of observation time spent on this source. However, for sources HerBS-83 and -150, we find several possible spectral lines (S/N $\sim$ 3), with inconclusive redshifts associated with them. This could point to several sources in a line-of-sight (LOS) alignment blended into a single sub-mm source in the \textit{Herschel} selection image. A high fraction of bright \textit{Herschel} sources are gravitationally lensed (e.g. \citealt{neg10}), potentially causing us to observe the foreground, lower-redshift source together with the background source.
Theoretically, \cite{hay13} found the possibility for LOS blending to be larger than the possibility of spatially associated blending ($\Delta z < 0.02$), though this is a source of controversy. Observationally, in one extreme case, \cite{zav15} found one of their sources to actually be three LOS blended sources, which could be a similar situation to HerBS-83. However, more follow-up on these three sources are needed, both in band E0 and E1.

\subsection{Current and future high-redshift searches}
\label{sec:CurrentFutureZspec}
The combination of the large collecting area of the IRAM 30~m telescope and the real-time data analysis enables us to find a CO line within around ten hours of observing time and subsequently confirm the redshift via the detection of a second line for these extremely luminous, mostly lensed sources. Higher-J (J$_{\rm up}$ $\geq$ 6) CO-lines are necessary for the robust redshift determination of high-redshift sources (z $>$ 3.5) within the 2 and 3~mm windows. Their line luminosities depend strongly on the internal properties of the galaxies, making these luminosities unpredictable. This can cause follow-up programs to suffer from biases towards sources with bright high-J CO-lines, which is indicative of thermalised systems and AGN.

Identifying spectroscopic redshifts with atomic spectral lines could help us, as [CII], [OI], [OIII] and [NII] are typically more luminous than the CO-lines. These atomic spectral lines are typically at slightly shorter wavelengths (e.g. $\lambda$ $<$ 1\,mm for [CII] at $z<5.3$). ALMA (e.g. \citealt{Lap17} and \citealt{Tam19}) and the \textit{Herschel} SPIRE FTS \citep{geo13,zha18} have demonstrated the viability of finding redshifts with atomic lines at high redshift, and new instruments are under development which will probe this regime with enough bandwidth to increase the speed of redshift searches. The upgrade to ZEUS, ZEUS-2, has already demonstrated a significant increase in the bandwidth \citep{fer14,vis18}. The relatively new technology, so-called Microwave Kinetic Inductance Detectors (M-KIDs --- \citealt{yat11}), promises a similar increase in bandwidth, with the added versatility, and a significant decrease in instrument size and complexity. Instruments such as DESHIMA \citep{end12, end19} and MOSAIC \citep{bas17} may allow 10m-class telescopes to compete with ALMA in redshift searches. This evolution in detector technology comes in a time when new telescopes are envisaged, such as the 50m-class telescope AtLAST/LST \citep{kaw16}, CCAT-prime and the SPICA space mission. To summarize, this promises a future where entire sub-mm samples can be followed-up with sub-mm spectroscopy, offering a less biased view of extreme star-forming galaxies. 

Redshift searches with interferometers such as ALMA and NOEMA, have also been conducted (e.g. \citealt{wei13,fud17,ner20}) and are currently ongoing (e.g. $z$-GAL\footnote{\url{http://www.iram.fr/~z-gal/}}, PIs: Cox, Bakx and Dannerbauer). These telescopes achieve reliable spectral line detections ($\sigma > 10$) {\color{referee} of lensed SMGs} with only very short on-source integration times (several minutes). This sensitivity is due to the large combined collecting area of the array, and also due to their smaller collective beam size, {\color{referee} which is similar to the typical size of the target sources ($\sim 1 - 2$\arcsec). This increases the sensitivity of the observation, since most emission of the source is contained within one beam, while collecting fewer photons from sources of noise, such as the atmosphere or the cosmic infrared background.}
These deep observations can negate the effect of biases, by ensuring even relatively faint high-J CO lines are observed.
The high demand for these institutes compared to the small availability of observation time requires optimised tuning set-ups, which can be found by maximising towards multiple-line detections (\textit{blue fill} in Figure \ref{fig:search}) within the redshift region one expects the sources to be.

\section{Physical properties}
\label{sec:discussion}
In Section \ref{sec:dustTemp}, we use the spectroscopic redshift to calculate the rest-frame wavelength of the photometry, and use it to derive dust temperatures and dust masses. In Section \ref{sec:molMass}, we calculate the luminosity of the spectral lines, in order to calculate the molecular gas masses. We compare the molecular gas masses to the dust mass and star formation rate in Section \ref{sec:G2D}. In Section \ref{sec:COSLED}, we compare the CO spectral line energy distributions (SLEDs). We briefly look at the effect of gravitational lensing on these SLEDs in Section \ref{sec:diffLens}. Finally, we investigate the gravitational lensing of the sources in Section \ref{sec:areoursourceslensed}.

\begin{table*}
\caption{Physical properties based on detected spectral lines}
\label{tab:specprop}
\resizebox{\textwidth}{!}{\begin{tabular}{lcccccc} \hline\hline
Source     & Transition & $\mu$L$'_{\rm spec line}$ & $\mu$L$'_{\rm{CO(1-0)}}$ & $\mu$M$_{\rm{mol}}$  &  t$_{\rm{dep}}$ &  $\delta$ \\ 
 & & [10$^{10}$ K km/s pc$^{2}$] & [10$^{10}$ K km/s pc$^{2}$]  & [10$^{10}$ M$_{\odot}$] & [Myr] & \\ \hline 
HerBS-52   & CO(1-0)   & 20.1 $\pm$ 3.1         & 20.1 $\pm$  3.1       & 16.1 $\pm$ 2.4  & 24  & 60 \\
            & CO(4-3)  & 19.7 $\pm$ 2.3         & 48.0 $\pm$  5.7       & 38.4 $\pm$ 4.6  & 57  & 142 \\
            & CO(5-4)  & 13.5 $\pm$ 1.8         & 42.3 $\pm$  5.6       & 33.8 $\pm$ 4.5  & 50  & 125 \\
            & CO(6-5)  & 9.0 $\pm$ 2.0        & 42.7 $\pm$  9.5       & 34.1 $\pm$ 7.6  & 51  & 126 \\
\hline
HerBS-61   & CO(4-3)   & 20.8 $\pm$ 2.9         & 50.8 $\pm$  7.1       & 40.7 $\pm$ 5.7  & 72  & 214 \\
            &  [C\,{\sc i}](1-0) & $<$ 10.7                     &           & $<$160 & $<$288 \\
            & CO(6-5)  & 3.7 $\pm$ 0.9       & 17.5 $\pm$  4.4       & 14.0 $\pm$ 3.5  & 25  & 74 \\
\hline
HerBS-64   & CO(1-0)   & 11.9 $\pm$ 2.3         & 11.9 $\pm$  2.3       & 9.51 $\pm$ 1.83  & 12 & 35 \\
            &  CO(2-1) & 8.1 $\pm$ 1.2        & 9.68 $\pm$  1.4       & 7.75 $\pm$ 1.12  & 10 & 29 \\
            & CO(3-2)  & 6.3 $\pm$ 2.2        & 12.1 $\pm$  4.3       & 9.68 $\pm$ 3.45  & 13 & 36 \\
            & CO(4-3)  & 8.6 $\pm$ 2.0        & 21.0 $\pm$  5.0       & 16.8 $\pm$ 4.0    & 22  & 62 \\
            & CO(6-5)  & 4.5 $\pm$ 1.0        & 21.3 $\pm$  4.6       & 17.0 $\pm$ 3.7  & 22  & 63\\
            & H$_{\rm 2}$O 2$_{\rm 11}$-2$_{\rm 02}$                          & 2.5 $\pm$ 0.8       &    \\
            & CO(7-6)  & $<$ 2.1                     & $<$ 11.7              & $<$ 9.36              & $<$ 12          & $<$ 35  \\
            & [C\,{\sc i}](2-1)  & $<$ 2.1                     & & $<$ 21 & $<$27 \\
\hline
HerBS-177    & CO(4-3) & 17.7 $\pm$ 3.8         & 43.1 $\pm$  9.2       & 34.5 $\pm$ 7.4  & 51 & 119 \\
            & [C\,{\sc i}](1-0)  & 7.1 $\pm$ 1.9        &  & 106 $\pm$ 29 & 159  \\
            & CO(6-5)  & 12.7 $\pm$ 1.2         & 60.6 $\pm$  5.5       & 48.5 $\pm$ 4.4  & 72 & 167 \\
\hline
\end{tabular}}
\raggedright \justify \vspace{-0.2cm}
\textbf{Notes:}  Col.~(1): Source name. Col.~(2): CO transition. Col.~(3): Line luminosity without magnification correction. Col.~(4): Predicted CO(1-0) line luminosity according to the average CO spectral line energy distribution based on \cite{bot13}. Col.~(5): Molecular hydrogen mass in the galaxy, derived from $\alpha$ = 0.8. Col.6): Depletion time is defined as the molecular gas mass divided by the star formation rate found in Table \ref{tab:photprop}. Col. (7): Gas-to-dust ratio, $\delta$, is the molecular gas mass divided by the dust mass.
\end{table*}

\subsection{Dust temperatures and mass estimates}
\label{sec:dustTemp}
\begin{figure}
\includegraphics[width=\linewidth]{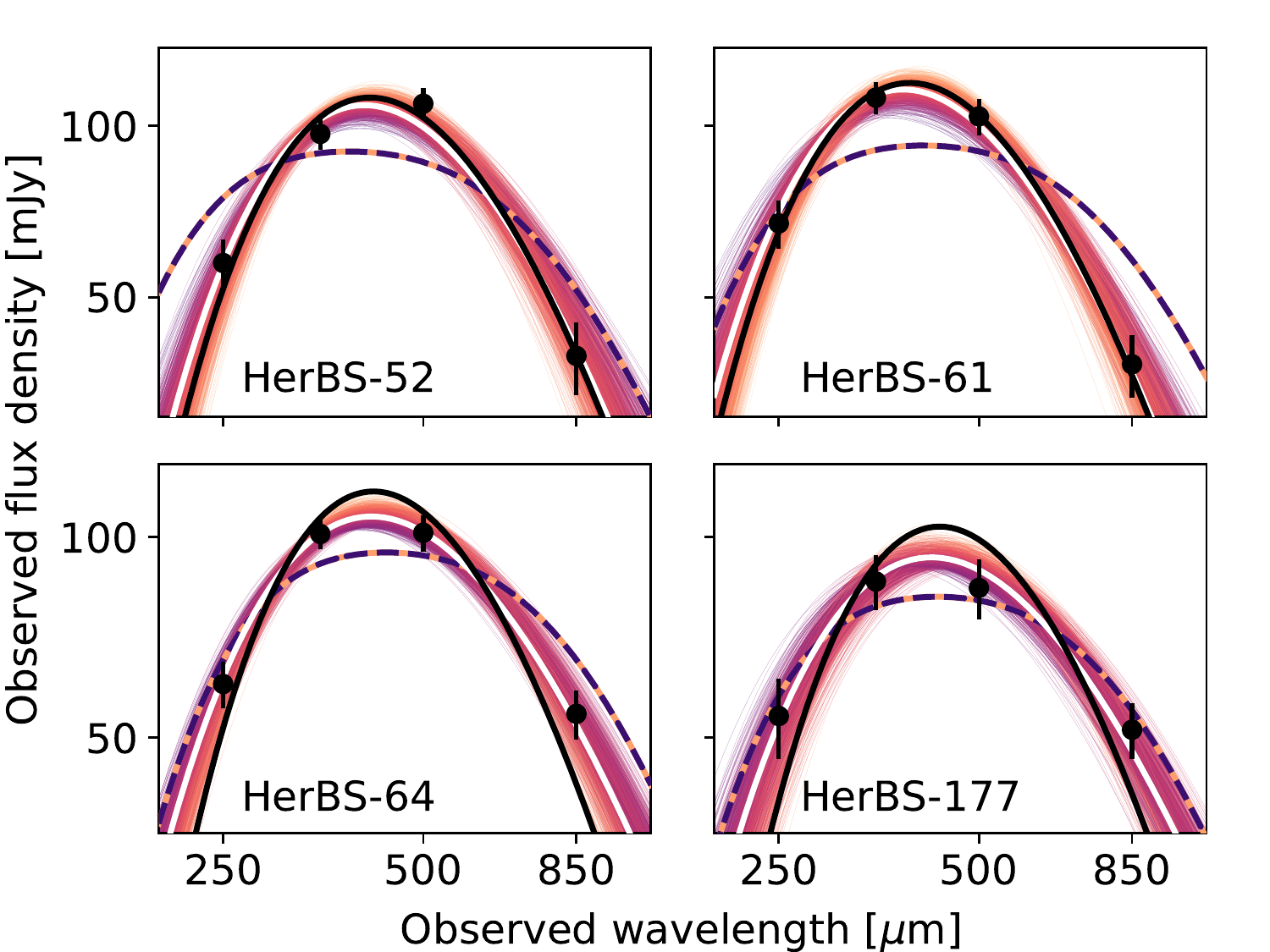}
\caption{We fit the continuum fluxes (\textit{black points}) for all sources with spectroscopic redshift for various spectral templates. The \textit{dashed line} shows the best-fit template from \citet{bak18}, which is derived by fitting a two-temperature modified black-body to \textit{Herschel}/SPIRE and JCMT/SCUBA-2 photometry of 24 HerBS sources with $z_{\rm spec}>1.5$. The \textit{black solid line} shows the best-fit of a single-temperature modified black-body assuming a dust-emissivity coefficient ($\beta$) of 2. We also show the best-fit spectra of a single-temperature modified black-body, with a variable $\beta$-value, for 1000 iterations of the Monte-Carlo code (\textit{purple-to-orange thin lines}), where the \textit{white line} shows the most likely solution. }
\label{fig:dustTemperature}
\end{figure}
We use the spectroscopic redshift to calculate the rest-frame wavelength of the photometry of each HerBS source. We use this photometry to fit a dust temperature to our sources, which we show in Figure \ref{fig:dustTemperature}. We fit the photometry for using two different modified black-body templates, for one template, we allow $\beta$ to vary, for the other template, we fix $\beta$ to 2. Here, we use the equations from \cite{dac13} and \cite{zha16} to include both the effect of dust heated by the CMB and the effect of the contrast against the CMB emission.

We find dust temperatures between 30 and 35~K, typical for SMGs at z~$\sim$~3.5 (e.g. \citealt{jin19}), which we show for each source in Table \ref{tab:specredshifts}. We find similar bolometric luminosities for our sources as detailed in Table \ref{tab:photprop}, calculated by integrating the template from \cite{bak18} from 8 to 1000~$\mu$m rest-frame. {\color{referee}We use this template, since it is based on the photometry of similarly-selected sources -- detailed in \cite{bak18}.} We determine the average dust temperature of the sources by fitting a single-temperature modified blackbody, with a $\beta$ of 2.0, to the galaxy.

We use the estimated dust temperature to calculate a total dust mass using 
\begin{equation}
M_{d} = \frac{S_{\nu}D_{\rm L}^2}{(1+z)\kappa_{\rm rest}B_{\nu}(\lambda_{\rm rest},T_d)}.
\end{equation}
S$_{\nu}$ is the observed flux density, $D_{\rm L}$ is the luminosity distance, $z$ is the spectroscopic redshift, $\kappa_{\rm rest}$ is derived by interpolating the values from \cite{dra03}, and $B_{\nu}(\lambda_{\rm rest},T_d)$ is the black-body radiation expected for a $T_d$ [K] source at $\lambda_{\rm rest}$. We use the 850 $\mu$m fluxes, as they probe the cold dust component of the spectrum.

We find dust masses of the order $2\times10^{9}$~M${\odot}$, which are similar to the dust masses found for other high-redshift sources selected from wide-area surveys, such as those in \cite{har16,fud17} and \cite{yan17}.

\subsection{Molecular gas masses}
\label{sec:molMass}
We detail the line properties in Table \ref{tab:specprop}. We calculate the line luminosity with the equation from \cite{sol97}:
\begin{equation}
L' =3.25\times10^7 S_{\rm CO}\Delta v f_{\rm obs}^{-2} D_{\rm L}^2(1+z)^{-3}.
\end{equation}
with the integrated flux, $S_{\rm CO}\Delta v$, in Jy km s$^{-1}$, the observed frequency, $f_{\rm obs}$, in GHz, and the luminosity distance, $D_{\rm L}$, in Mpc. 

We use the line luminosity ratios from \citet{ivi11} and \citet{bot13} to calculate the CO(1-0) line luminosity. The molecular hydrogen mass in the galaxy is calculated using
\begin{equation}
M_{\rm H_2} = \alpha L'_{\rm CO(1-0)}.
\end{equation}
We adopt $\alpha$ = 0.8 $M_{\odot}$/(K km/s pc$^2$). Similar to \cite{har16,fud17} and \cite{yan17}, we find molecular gas masses on the order of 10$^{11}$ M$_{\odot}$, not corrected for magnification.

Here, we note the factor of two difference between the CO(1-0)-derived molecular mass for HerBS-52 and -64 compared to the masses derived from the other transitions. The ratios tabulated by \cite{bot13} over-predict the CO(1-0) luminosity, suggesting that HerBS-52 and -64 are more thermalised than the average source in \cite{bot13}. Since the CO(1-0) fluxes give the most accurate description of the CO-derived molecular gas masses, we note the need for direct CO(1-0) observations to derive accurate molecular gas masses. Moreover, even for HerBS-61 where we only have J~$>3$ CO lines, we find a factor of 3 difference in molecular gas mass between the two lines.

We also derive the molecular gas masses from the [C\,{\sc i}](1-0) and [C\,{\sc i}](2-1) upper limits and the single [C\,{\sc i}](1-0) detection. [C\,{\sc i}] emission traces the molecular gas more reliably (e.g. \citealt{pap04}), however is typically fainter than CO emission. We use the equations from \cite{wei05} and \cite{wal11} to calculate the [C\,{\sc i}] mass from both the [C\,{\sc i}](1-0) and [C\,{\sc i}](2-1) line luminosity estimates. Since no sources have both [C\,{\sc i}](2-1) and [C\,{\sc i}](1-0) observations, we assume that the [C\,{\sc i}] excitation temperature is equal to the dust temperature. From \cite{wal11}, we adopt the average carbon abundance X[C\,{\sc i}]/X[H$_2$] = 8.4 $\times$ 10$^{-4}$. For both the upper limits of HerBS-61 and -64, we find molecular gas masses agreeing with the CO-derived molecular gas masses. For HerBS-177, where we have a [C\,{\sc i}] line detection, we find double or triple the amount derived from the CO emission. This could be due to large amounts of cold, sub-thermally ionised gas, which are not observable with J~$>3$ CO emission \citep{pap04, pap18}. The spread in molecular gas mass estimates indicates the need for high-redshift J~$<4$ CO and [C\,{\sc i}] observations, {\color{referee}although we note the low significance of the [CI](1-0) observation, and the lack of the [CI](2-1) line, which is crucial for the correct excitation temperature. Moreover, the estimated CO-dark gas could change depending on differential lensing effects.}

\subsection{Gas depletion timescale and gas-to-dust ratio}
\label{sec:G2D}
Gas depletion times, defined as the molecular gas mass divided by the star formation rate (t$_{\rm dep}=\mu$M$_{\rm mol}$/$\mu$SFR), are on the order of tens of millions of years. It is important to note that the star-formation rates are derived from the photometry, which are very dependent on the assumed initial mass function, and can vary significantly (e.g. \citealt{tac20}). These suggest that these sources are undergoing very short-lived bursts in star formation, or have to be accreting significant amounts of intergalactic gas to sustain this rate of star formation. In the case of HerBS-177, the [C\,{\sc i}]-derived molecular gas mass indicates that we are missing the CO-dark molecular gas in our estimates, which could also fuel the star formation and sustain the starburst phase over longer timescales. These time-scales are similar to those found for other high-redshift sources selected from wide-area surveys, such as those in \cite{har16,fud17} and \cite{yan17}.

The gas-to-dust ratio, $\delta$, is the molecular gas mass divided by the dust mass. We find ratios of the order tens to hundreds. This is similar to what is seen for local galaxies and high-redshift SMGs (e.g. \citealt{dun18}).

\subsection{CO spectral line energy distribution}
\label{sec:COSLED}
\begin{figure}
\includegraphics[width=\linewidth]{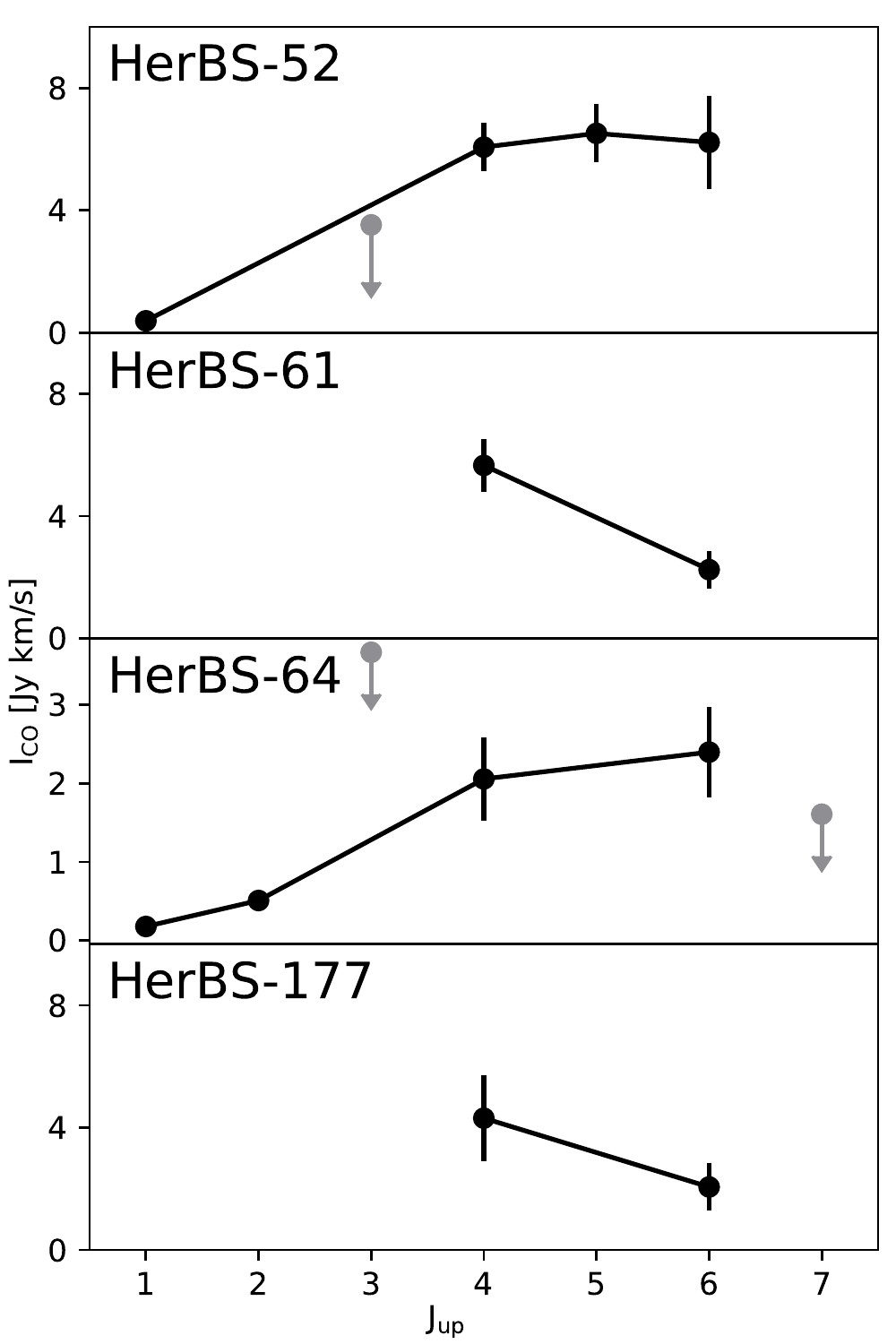}
\caption{The CO SLED for the four sources from our sample with spectroscopic redshifts. The velocity-integrated flux I$_{\rm CO_{J_{up},J{up-1}}}$ for each CO transition, J$_{\rm up}$, shows the diverse behaviour of the CO spectral line energy distributions within our sample. Arrows indicate upper limits, where the integrated flux is taken to be less than three times the off-line standard deviation, assuming a FWHM of 500 km~$s^{-1}$.}
\label{fig:jladder}
\end{figure}

Figure~\ref{fig:jladder} shows the velocity-integrated flux I$_{\rm CO}$ of each CO transition of the sources with identified J-transitions. The CO-ladders of HerBS-52, and -64 appear to peak around $J=5$ or 6, whilst both HerBS-61 and -177 appear to have already sloped downward {\color{referee} at or} below $J=4$. 

\begin{figure}
\includegraphics[width=\linewidth]{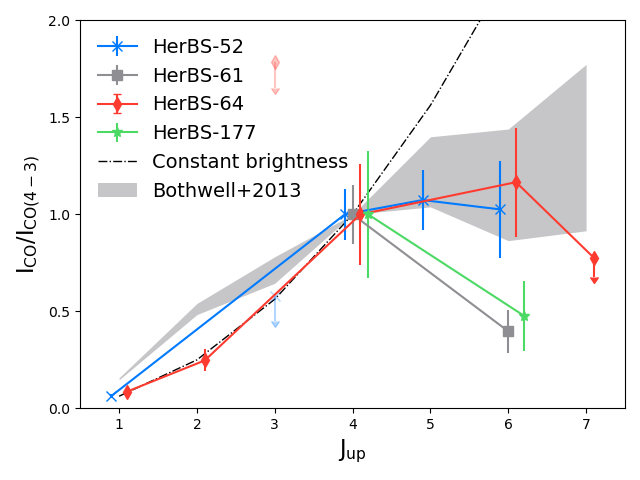}
\caption{Comparison of the individual CO line luminosity ratios, normalised against CO(4-3), for different J$_{\rm up}$. For comparison, we plot against two SLEDS; fully thermalised SLED, and the average SLED of \citet{bot13}.}
\label{fig:COladder}
\end{figure}

Figure~\ref{fig:COladder} shows the integrated flux of each detected CO line of the sources with a spectroscopic redshift. We normalize the CO spectral line energy distribution (CO SLED) to the CO(4-3), as this transition was detected for all sources. We compare the CO SLEDs against two types of SLED profiles. Firstly, the constant brightness profile indicates the theoretical maximum emission in each CO transition, in the case where all excitation states are equally populated. 
Secondly, we compare against the line ratios of non-lensed SMGs from \cite{bot13} based on CO(1-0) observations by \cite{rie11} and \cite{ivi11}, with S$_{\rm 850\mu m}$ flux densities about 6 to 20 times lower than the sources in our sample.

The slope in Figure \ref{fig:jladder} and the comparison in Figure \ref{fig:COladder} show that HerBS-52 and HerBS-64 are more thermalised system (e.g. \citealt{wei05,wei07,wer10}). This is contrasted by HerBS-61 and HerBS-177, which appear to already slope downward, suggestive of a less-thermalised system.
We note that the CO(3-2) emission of HerBS-52 suggests a lower flux than expected from a constant-brightness emission, suggesting the upper-limit on the CO(3-2) flux is incorrect due to the poor quality baselines.

We derive the line luminosity ratios, $r_{\rm J_a, J_b}$, for each of our sources, using equation
\begin{equation}
r_{\rm J_a, J_b} = \frac{L'_{a}}{L'_{b}} = \frac{I_{a}}{I_{b}} \frac{J_{b}^2}{J_{a}^2}.
\label{eq:JaJb} 
\end{equation}
Here, $L'_{a}$ and $L'_{b}$ refers to the line luminosity of lines $a$ and $b$, $I_{a}$ and $I_{b}$ refers to the integrated line flux, and $J_a$ and $J_b$ refer to the upper quantum transition, commonly referred to as $J_{\rm up}$.

\begin{figure}
\includegraphics[width=\linewidth]{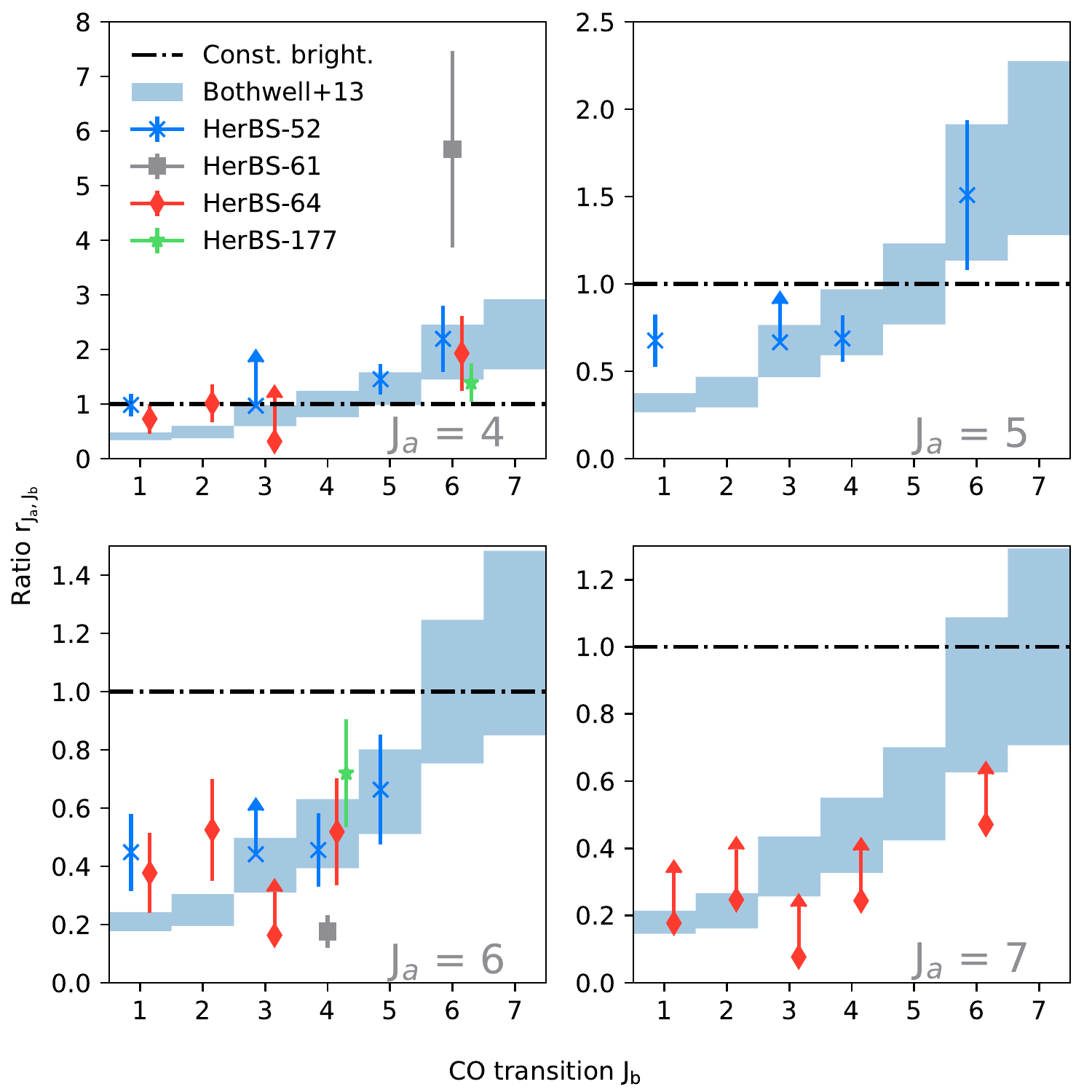}
\caption{Line luminosity ratios between the different lines are calculated based on the CO-line luminosities using equation \ref{eq:JaJb}. We compare the ratios to the expected ratios of the compilation of \citet{bot13} (\textit{blue fill}), and the constant brightness profile (\textit{dashed line}). The constant brightness profile assumes equal excitation among all states, and thus is 1.0 for all transitions.}
\label{fig:ratios}
\end{figure}

For each of our sources, we plot the ratios in Figure~\ref{fig:ratios}. We compare the sample to the ratio from \cite{bot13}, and the constant brightness SLED profile. 
In general, the fluxes agree with the \cite{bot13} ratios. The only major outlier is the CO(6-5) flux of HerBS-61, which is significantly fainter than predicted, resulting in a high data point. Notably, the lower-J fluxes of HerBS-52 suggest that these transitions are fully thermalised, while the higher transitions are not. We note that not all source ratios lie within the predicted values of \cite{bot13}, suggesting a large spread in the internal conditions of the ISM.

\subsection{Differential lensing}
\label{sec:diffLens}
The magnification factor of a strongly-lensed source is dependent on the position of the emission in the source plane (e.g. \citealt{ser12}). Since emission at a certain CO transition is determined by the ISM conditions, a different magnification factor between various ISM conditions can skew the analysis of the galaxy, especially for unresolved observations such as these IRAM 30m and GBT observations. Since we suspect our sources are lensed from both optical and near-infrared follow-up \citep{neg17,bak20}, and from the following Section \ref{sec:areoursourceslensed}, we investigate the potential effects of this on our SLED analysis.

Among our sources, both HerBS-52 and HerBS-61 have smaller line widths (FWHM) for their CO(6-5) line, when compared to their lower-J transitions. HerBS-177 actually has a larger velocity width for its CO(6-5) line than the other two spectral lines, although the significance of this velocity difference is smaller. HerBS-64 does not appear to have a large difference in line width, where all lines are within 350 to 450~km/s. Discrepancies among the line widths could suggest differential lensing, although these discrepancies can also be explained by a difference in ISM distribution.

The CO luminosity ratios in Figure \ref{fig:ratios} provide another indication of differential lensing. The CO(4-3) over CO(6-5) ratio ($r_{\rm 4,6}$) of HerBS-61 is around 1.5$\sigma$ larger than expected from \cite{bot13}. Similarly, both HerBS-52 and HerBS-64 have a higher ratio than expected for the CO(6-5) over CO(1-0) -- and over CO(2-1) for HerBS-64 -- luminosity ratio ($r_{\rm 6, 1}$ and $r_{\rm 6, 2}$). 
Unfortunately, there exists a large degeneracy between the internal ISM conditions and the effects of differential lensing. This is further exacerbated by the low S/N of the spectral lines, which stresses the need for resolved observations.

\subsection{Are our sources lensed?}
\label{sec:areoursourceslensed}
{\color{referee}Analogous to the Tully-Fisher relationship \citep{tul77}, we compare the CO line width against the line luminosity to find evidence for the lensing nature of our sources \citep[e.g.][]{har12,bot13,ara16,dan17,ner20}. Lensing amplifies the apparent CO luminosity, while the line width remains unaffected.} \cite{har12} showed that this diagnostic could be used to show that the brightest {\it Herschel} galaxies are mostly lensed, {\color{referee} with relatively large CO luminosities compared to their velocity widths}. In our case, we know that there is a large likelihood of our sources being lensed, as they are selected with 500$\mu$m fluxes around the lensing selection function by \cite{neg10} of $\sim$100 mJy. 
From previous work \citep[e.g,][]{ivi11,rie11,bot13}, we know there exists a significant spread in the conversion of CO(4-3) flux to CO(1-0), and we therefore examine the lensing properties of our sources using the CO(4-3) line, and compare them against sources with detected CO(4-3) fluxes. 

\begin{figure}
\includegraphics[width=\linewidth]{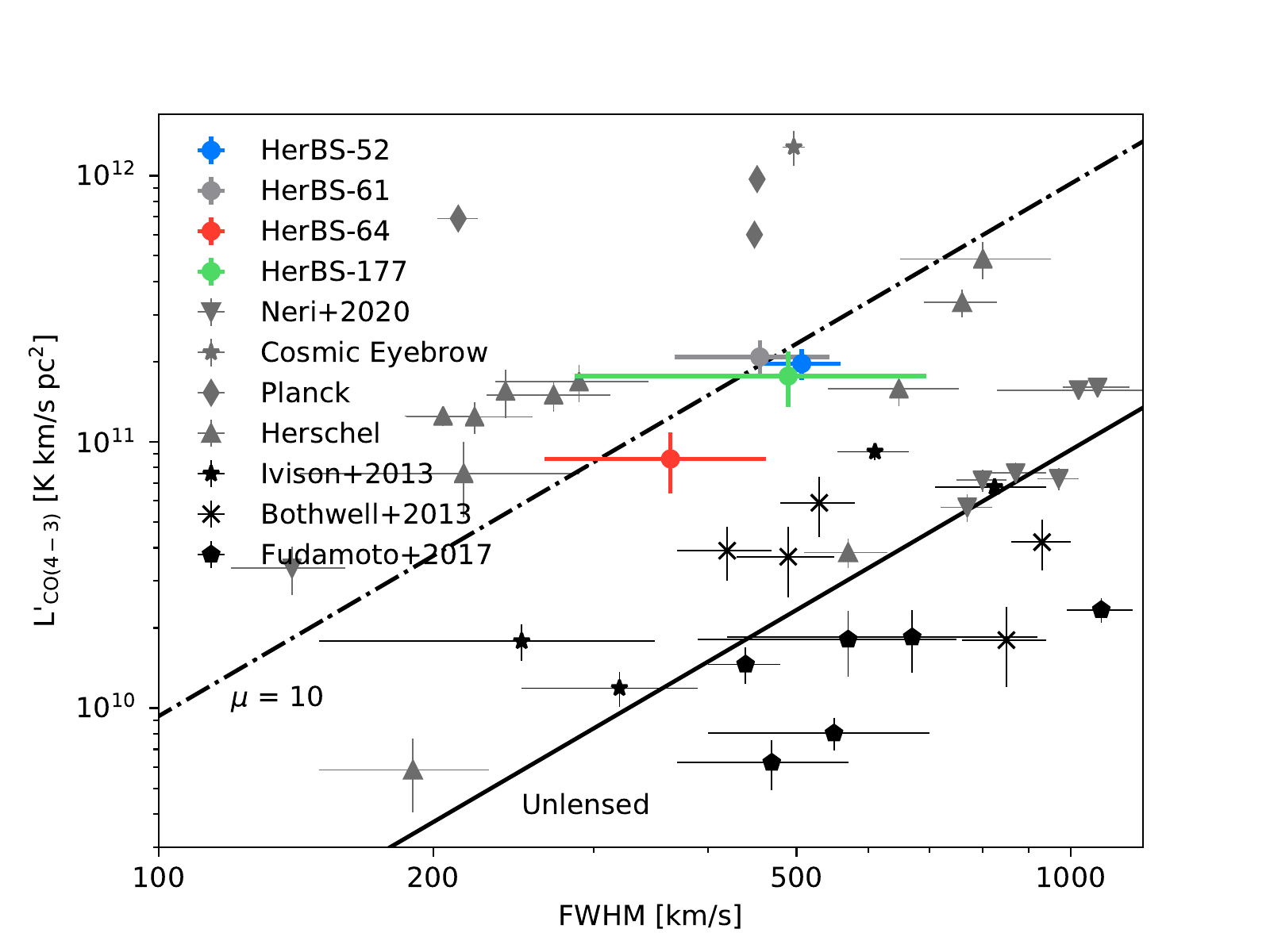}
\caption{The CO(4-3) line luminosity is shown against the line width for the sources with robust redshift identifications. Their bright line luminosity, compared to their line width, suggests they are gravitationally lensed. Unlensed sources (\textit{black symbols}) lie close to the solid line from \citet{bot13}, as can be seen for the (predominantly) unlensed sources from \citet{bot13}, \citet{ivi13} and \citet{fud17}. Lensed sources (\textit{dark grey symbols}), such as known lenses (Cosmic Eyebrow; \citealt{dan19}) and the ones selected from the {\it Planck} \citep{can15,har16} and \textit{Herschel} surveys \citep{cox11,zav15,yan17,ner20}. These lenses have more apparently-luminous spectral lines, while the gravitational lensing does not affect the line widths. Lensed sources are therefore expected in the upper-left part of this figure.}
\label{fig:FWHM}
\end{figure}

Figure~\ref{fig:FWHM} shows the CO(4-3) line luminosity of each source plotted versus the Full-Width at Half-Maximum (FWHM) of the line width. We compare against the CO(4-3) luminosity of other SMG samples: the sources from \cite{bot13}, the Planck sources from \cite{har16} and \cite{can15}, the brightest {\it Herschel} sources from \cite{yan17}, \cite{zav15} and \cite{geo13}, the Hyperluminous LIRG\footnote{This source was explicitly selected as a potential HyLIRG by its large linewidth} from \cite{ivi13}, the sources from \cite{ner20}, and the Cosmic Eyebrow \citep{dan19}. Finally, we compare against the ultra-red {\it Herschel} sources from \cite{fud17}, explicitly selected to contain a low fraction of lensed sources, aiming to find the most distant dusty starbursts \citep{ivi16}, although high-resolution ALMA imaging by \cite{ote17} suggests that $\sim$40\% of sources are gravitationally-lensed.

The unlensed relationship between the line luminosity and its velocity width is from \cite{bot13} (solid line). The dashed line, assuming a magnification factor of $\mu=10$, represents the magnified relation. We note a distinct split between lensed and unlensed sources. All our sources appear to agree with the $\mu=10$ line, and lie within the distribution of previously-observed lensed sources. This suggests that these sources are all magnified by the mass of foreground intervening galaxies.


\section{Conclusions}
\label{sec:conclusions}
In an effort to characterize the galaxies between the \textit{cosmic dawn} and \textit{cosmic noon}, we have targeted 7 bright \textit{Herschel} sources from the HerBS sample around redshift $z \sim 4$. For four of these sources, we have identified robust spectroscopic redshifts using the IRAM 30m telescope, with redshifts ranging between $z=3.4$ to 4.1. For two of these sources with spectroscopic redshifts, we obtained GBT observations of the J $<$ 4 CO lines. In this redshift search, the use of the 2~mm window proved indispensable to confirm the spectroscopic redshifts using the CO line transitions, as can be seen in our graphical analysis in Figure \ref{fig:search}. We therefore recommend using the 2~mm window together with the 3~mm window in future redshift searches. Both a lack of observation time and the potential of source confusion could explain the lack of a robust redshift for the three sources without a spectroscopic redshift identification.

The four sources with spectroscopic redshifts have dust temperatures around 30 to 35~K, and dust masses around 2~$\times$~10$^{9}$~M$_{\odot}$, similar to other sources selected with \textit{Planck}, \textit{Herschel} and the South Pole Telescope. We estimated the molecular gas mass in the galaxies using both low-J and high-J CO-lines and using the [C\,{\sc i}] line. We find molecular gas masses between 1 and 10~$\times$~10$^{11}$~M$_{\odot}$, however we note a large discrepancy between the molecular gas masses estimated using different spectral lines. There exists around a factor of 2 discrepancy between the CO(1-0) and J~$>3$ CO transitions, stressing the need for CO(1-0) lines to accurately estimate the molecular gas mass. The molecular gas estimate from atomic carbon ([C\,{\sc i}]) for one source is several times larger than from CO lines, suggesting the possibility that the majority is CO-dark gas.

The CO SLEDs of the four sources appear different, where HerBS-52 and HerBS-64 have increasing SLEDs, peaking around CO(5-4) or CO(6-5), while HerBS-61 and HerBS-177 appear to have downward-sloping SLEDs, suggesting a peak below CO(4-3). The behaviour of the first two sources is similar to the average behaviour seen in population studies, while the latter two appear slightly discrepant.

We find an indication of gravitational lensing for all four sources with spectroscopic redshifts. The luminosity of the CO line emission is larger than we expect from their line width. This indicates an amplification of the emitted spectral line due to gravitational lensing. {\color{referee}This is in line with what we expected, due to the presence of potential foreground galaxies identified in optical and near-infrared studies.}

\section*{Acknowledgements}
This work is based on observations carried out under project number 085-15, 076-16, 195-16 and 087-18 with the IRAM 30m telescope. IRAM is supported by INSU/CNRS (France), MPG (Germany) and IGN (Spain). We thank the anonymous referee for comments that helped us to improve our arguments and presentation in this paper. We would like to thank Claudia Marka and Pablo Torne for their (astronomical) expertise help during the IRAM observations, as well as comments on the paper's contents from Pierre Cox, Roberto Neri, Shouwen Jin, and Kevin Harrington. TB, MS and SAE have received funding from the European Union Seventh Framework Programme ([FP7/2007-2013] [FP\&/2007-2011]) under grant agreement No. 607254. TB further acknowledges funding from NAOJ ALMA Scientific Research Grant Numbers 2018-09B and JSPS KAKENHI No. 17H06130. The \textit{Herschel}-ATLAS is a project with \textit{Herschel}, which is an ESA space observatory with science instruments provided by European-led Principal Investigator consortia and with important participation from NASA. The \textit{Herschel}-ATLAS website is \url{http://www.h-atlas.org}. H.D. acknowledges financial support from the Spanish Ministry of Science, Innovation and Universities (MICIU) under the 2014 Ram\'on y Cajal program RYC-2014-15686 and AYA2017-84061-P, the later one co-financed by FEDER (European Regional Development Funds). JGN acknowledges financial support from the PGC 2018 project PGC2018-101948-B-I00 (MICINN, FEDER), the PAPI-19-EMERG-11 (Universidad de Oviedo) and the Spanish MINECO `Ramon y Cajal' fellowship (RYC-2013-13256). I.P.F. acknowledges support from the Spanish Ministerio de Ciencia, Innovaci\'on y Universidades (MICINN) under grant number ESP2017-86852-C4-2-R. M.J.M.~acknowledges the support of the National Science Centre, Poland through the SONATA BIS grant 2018/30/E/ST9/00208. This research made use of Astropy\footnote{http://www.astropy.org}, a community-developed core Python package for Astronomy \citep{ast13, ast18}.





\appendix

\section{Individual spectra}

\begin{figure*}
\includegraphics[width=\linewidth]{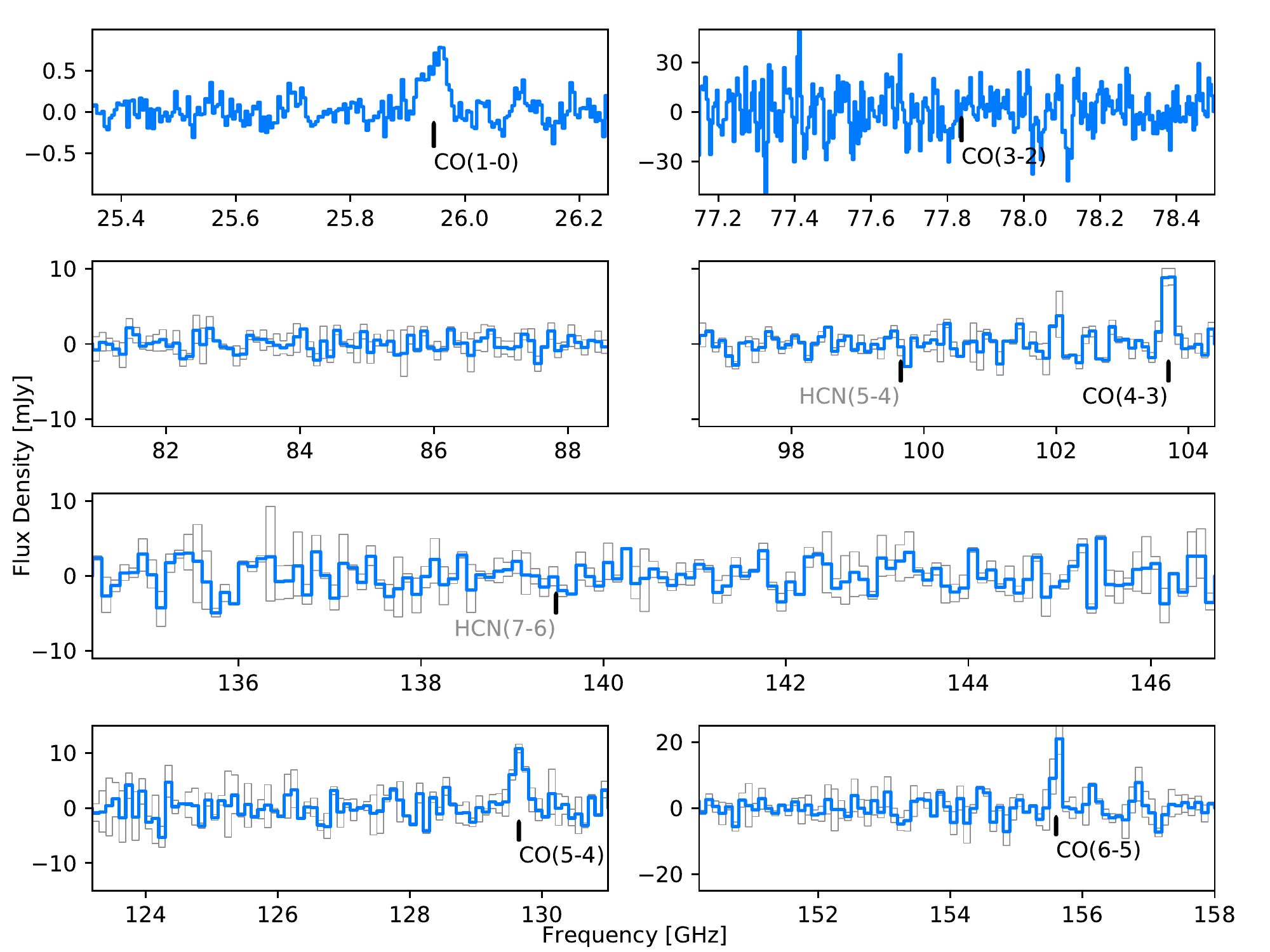}
\caption{The complete spectra taken of source HerBS-52. We show the spectra at 70~km/s for individual polarisations (\textit{grey}) and combined polarisations (\textit{blue}).}
\label{fig:HerBS52}
\end{figure*}

\begin{figure*}
\includegraphics[width=\linewidth]{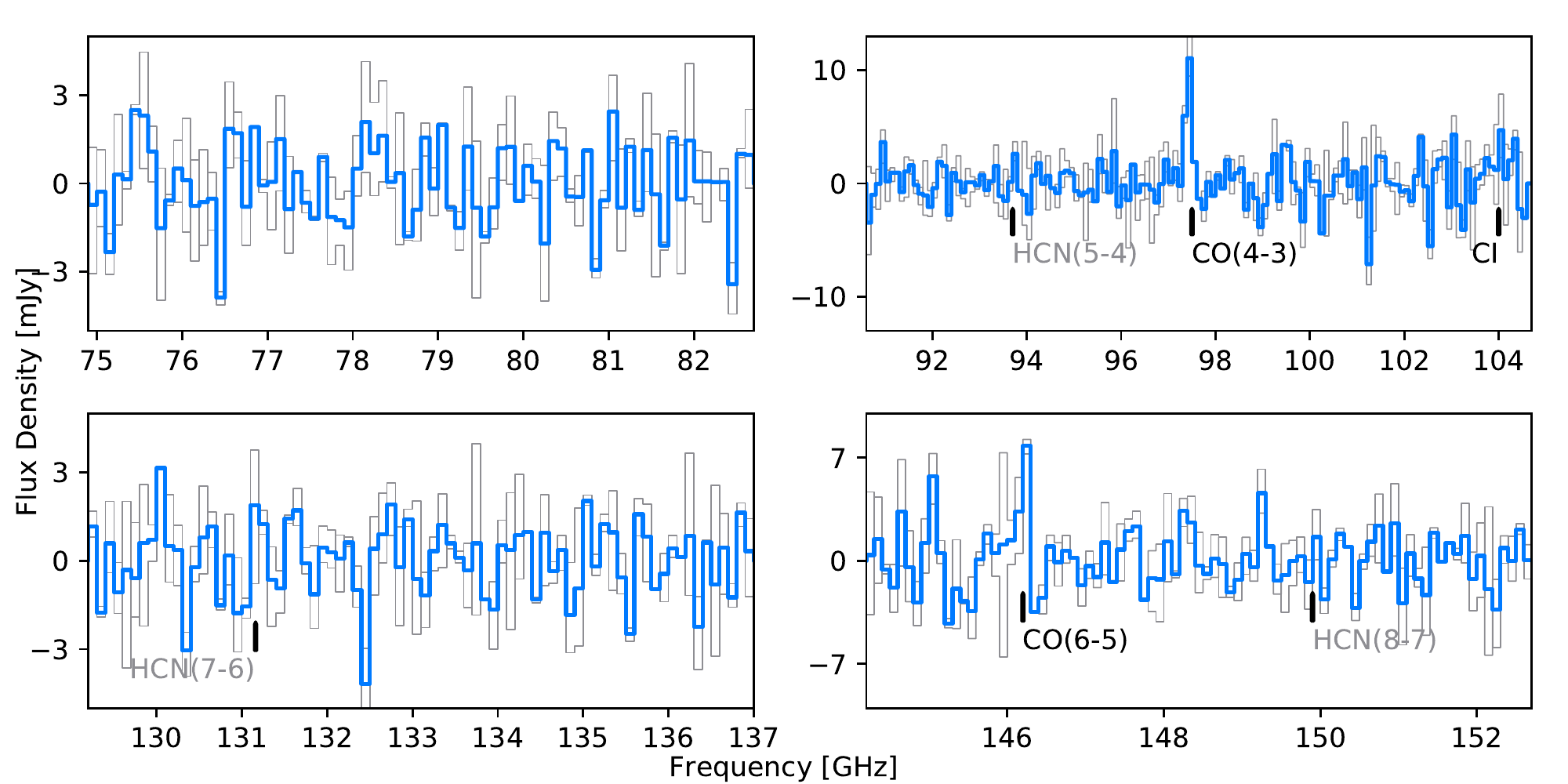}
\caption{The complete spectra taken of source HerBS-61. We show the spectra at 70~km/s for individual polarisations (\textit{grey}) and combined polarisations (\textit{blue}).}
\label{fig:HerBS61}
\end{figure*}

\begin{figure*}
\includegraphics[width=\linewidth]{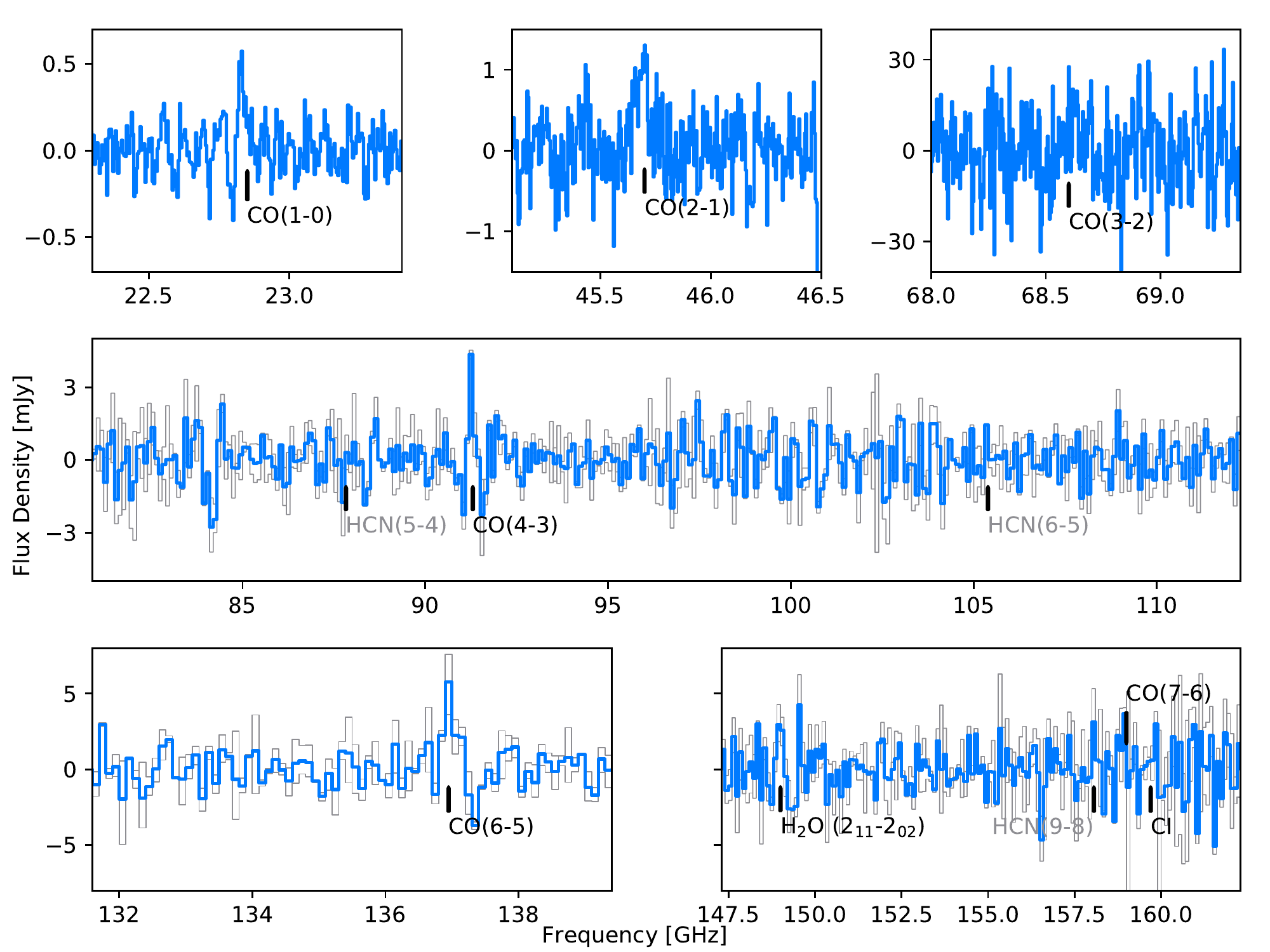}
\caption{The complete spectra taken of source HerBS-64. We show the spectra at 70~km/s for individual polarisations (\textit{grey}) and combined polarisations (\textit{blue}).}
\label{fig:HerBS64}
\end{figure*}

\begin{figure*}
\includegraphics[width=\linewidth]{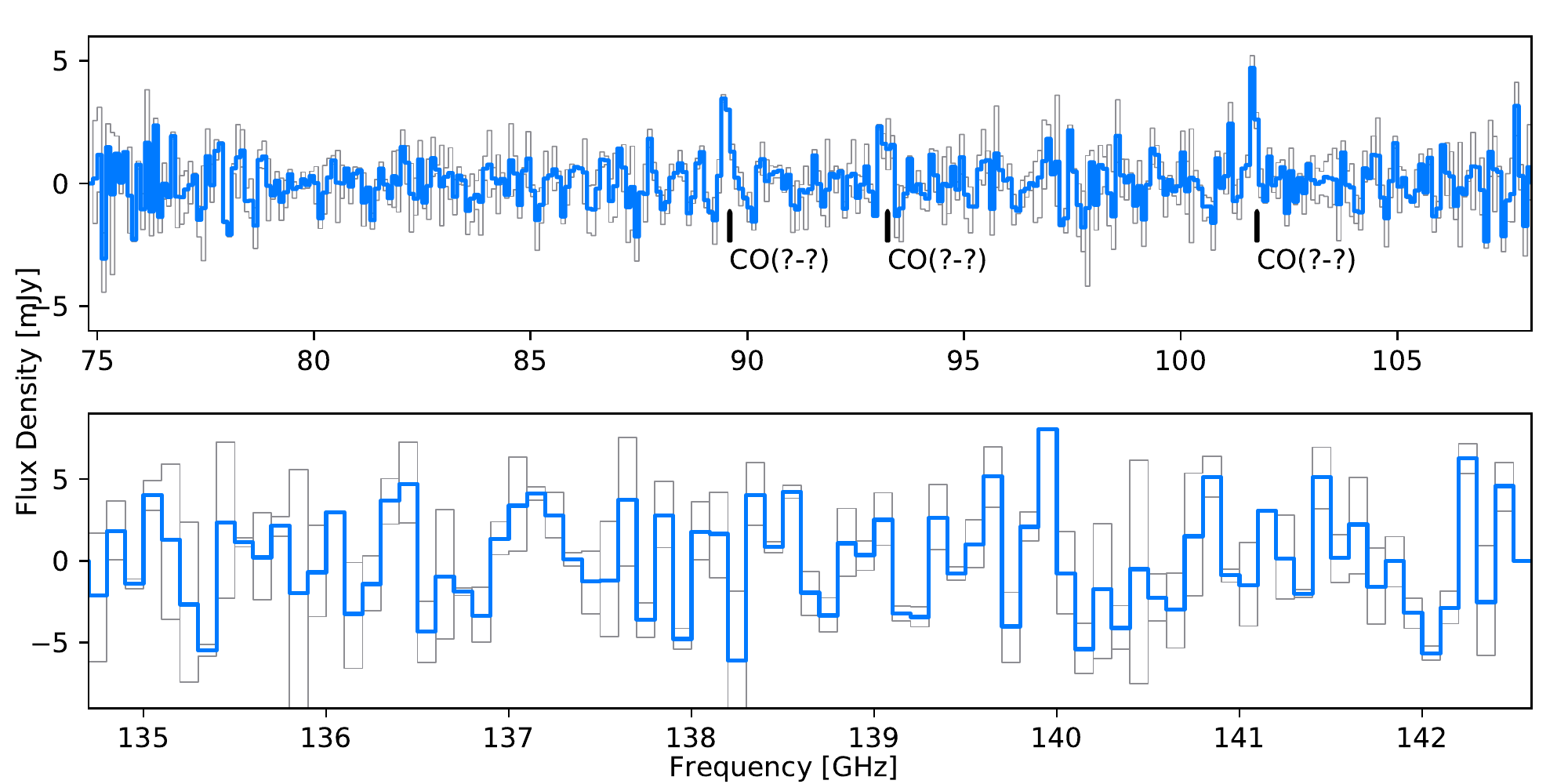}
\caption{The complete spectra taken of source HerBS-83. We show the spectra at 70~km/s for individual polarisations (\textit{grey}) and combined polarisations (\textit{blue}).}
\label{fig:HerBS83}
\end{figure*}

\begin{figure*}
\includegraphics[width=\linewidth]{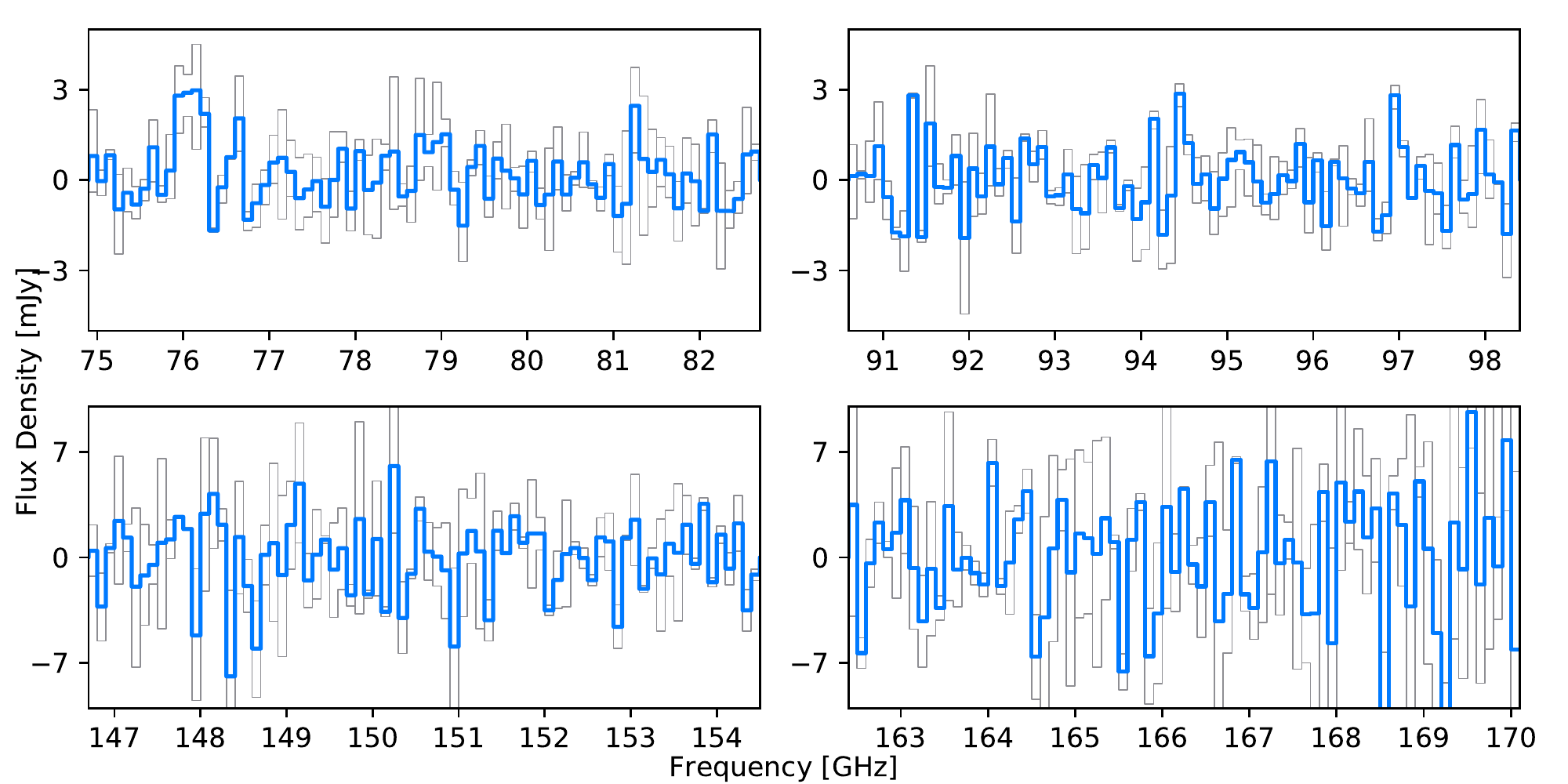}
\caption{The complete spectra taken of source HerBS-89. We show the spectra at 70~km/s for individual polarisations (\textit{grey}) and combined polarisations (\textit{blue}).}
\label{fig:HerBS89}
\end{figure*}

\begin{figure*}
\includegraphics[width=\linewidth]{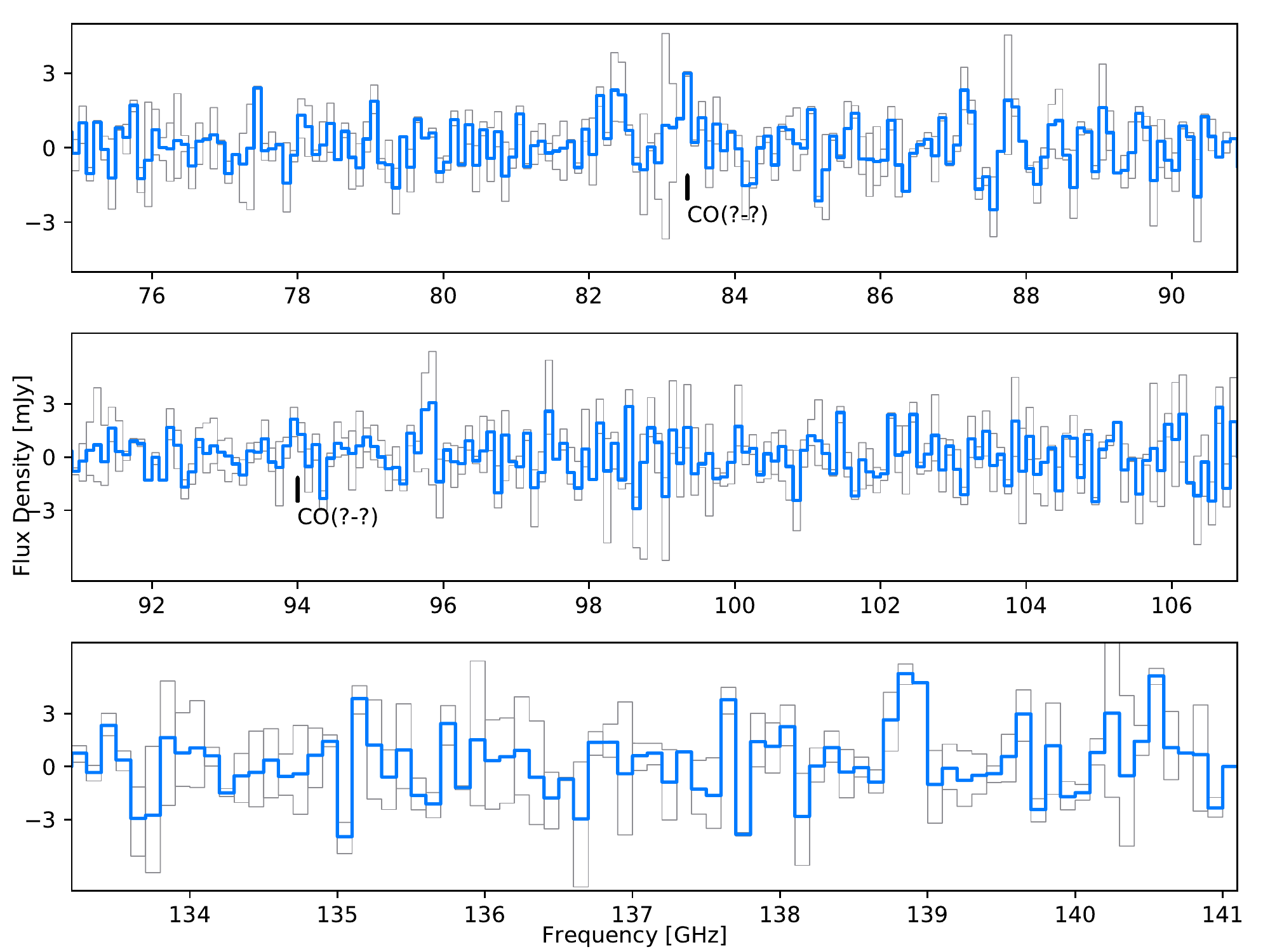}
\caption{The complete spectra taken of source HerBS-150. We show the spectra at 70~km/s for individual polarisations (\textit{grey}) and combined polarisations (\textit{blue}).}
\label{fig:HerBS150}
\end{figure*}

\begin{figure*}
\includegraphics[width=\linewidth]{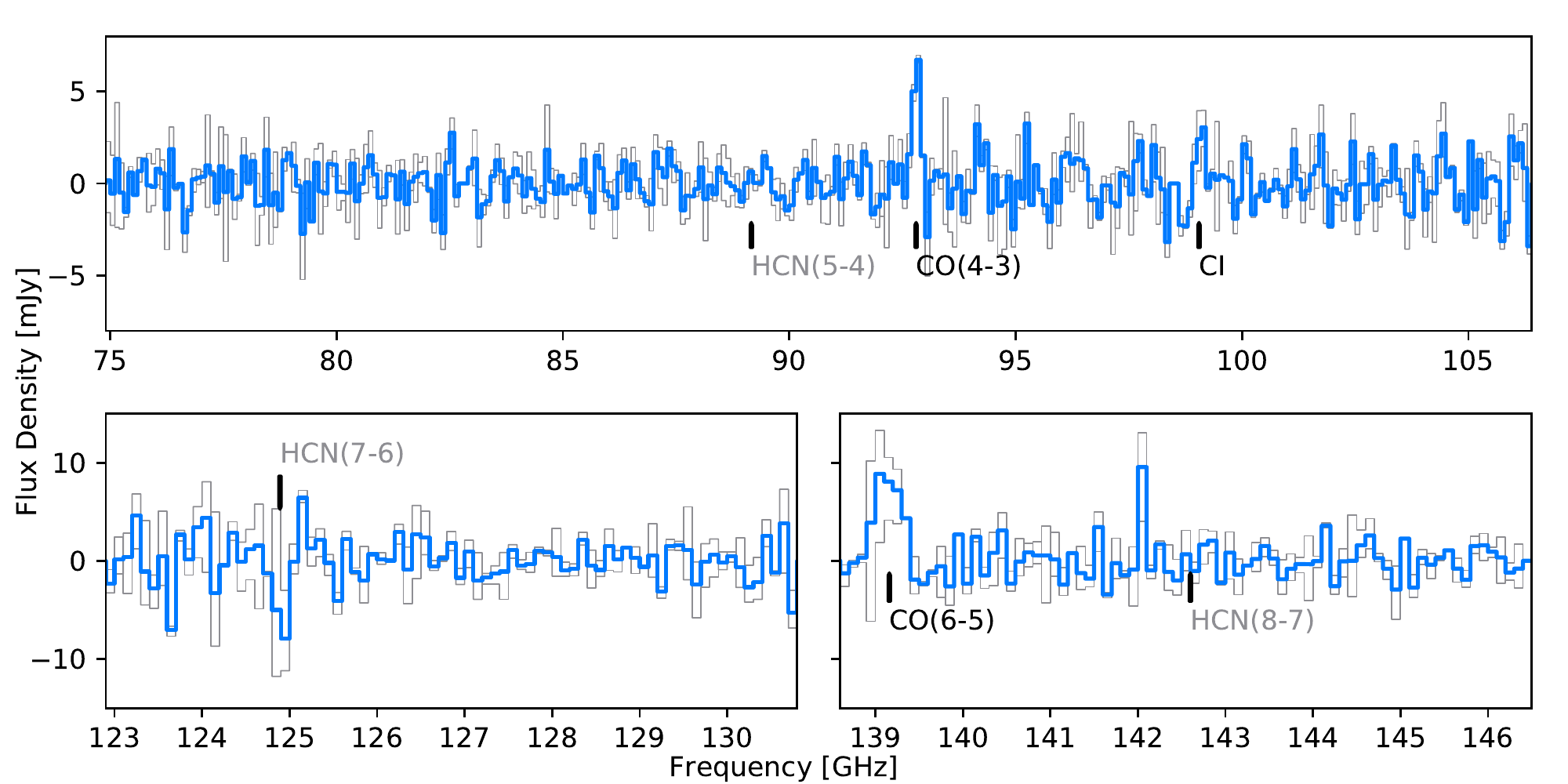}
\caption{The complete spectra taken of source HerBS-177. We show the spectra at 70~km/s for individual polarisations (\textit{grey}) and combined polarisations (\textit{blue}).}
\label{fig:HerBS177}
\end{figure*}




\bsp  
\label{lastpage}
\end{document}